%% file: main.tex
\documentclass[9pt, manuscript, screen, sigconf]{acmart}

\input{includes/packages}
\usepackage{framed}

\definecolor{shadecolor}{gray}{0.95}

\newenvironment{mybox}
{\begin{center}\begin{minipage}{0.98\columnwidth}%
  \MakeFramed{\advance\hsize-\width \FrameRestore}%
}
{\endMakeFramed\end{minipage}\end{center}}
 
\copyrightyear{2025}
\acmYear{2025}
\setcopyright{cc}
\setcctype{by}
\acmConference[ISCA '25]{Proceedings of the 52nd Annual International Symposium on Computer Architecture}{June 21--25, 2025}{Tokyo, Japan}
\acmBooktitle{Proceedings of the 52nd Annual International Symposium on Computer Architecture (ISCA '25), June 21--25, 2025, Tokyo, Japan}\acmDOI{10.1145/3695053.3730989}
\acmISBN{979-8-4007-1261-6/2025/06}

\begin{document}

\title[MicroScopiQ: Accelerating Foundational Models through Outlier-Aware Microscaling Quantization]{MicroScopiQ: Accelerating Foundational Models through Outlier-Aware Microscaling Quantization}

\author{Akshat Ramachandran}
\affiliation{%
  \institution{Georgia Institute of Technology}
  \country{Atlanta, GA}
}
\email{akshat.r@gatech.edu}

\author{Souvik Kundu}
\affiliation{%
  \institution{Intel Labs}
  \country{San Diego, CA}
}
\email{souvikk.kundu@intel.com}

\author{Tushar Krishna}
\affiliation{%
  \institution{Georgia Institute of Technology}
  \country{Atlanta, GA}
}
\email{tushar@ece.gatech.edu}
\renewcommand{\shortauthors}{Akshat Ramachandran, Souvik Kundu and Tushar Krishna}


\begin{abstract}
    Quantization of foundational models (FMs) is significantly more challenging than traditional DNNs due to the emergence of large magnitude values called outliers. Existing outlier-aware algorithm-architecture co-design techniques either use mixed-precision, retaining outliers at high precision but compromise hardware efficiency, or quantize inliers and outliers at the same precision, improving hardware efficiency at the cost of accuracy. To address this mutual exclusivity, we propose MicroScopiQ, a novel co-design technique that leverages pruning to complement outlier-aware quantization. MicroScopiQ retains outliers at higher precision while pruning a certain fraction of least important weights to distribute the additional outlier bits; ensuring high accuracy, aligned memory and hardware efficiency. We design a high-throughput, low overhead accelerator architecture composed of multi-precision INT processing elements and a network-on-chip called ReCoN that efficiently abstracts the complexity of supporting high-precision outliers. Additionally, unlike prior techniques, MicroScopiQ does not assume any locality of outlier weights, enabling applicability to a broad range of FMs. Extensive experiments across diverse quantization settings demonstrate that MicroScopiQ achieves state-of-the-art quantization accuracy, while delivering up to $\mathbf{3}\times$ faster inference and $\mathbf{2}\times$ lower energy consumption compared to existing alternatives. Code is available at: \href{https://github.com/georgia-tech-synergy-lab/MicroScopiQ-LLM-Quantization}{\texttt{MicroScopiQ-LLM-Quantization.git}}

\end{abstract}



\maketitle

  \input{sections/introduction}
  \input{sections/background_motivation}

  \input{sections/algo_methodology}
  \input{sections/hw_methodology}

  \input{sections/results}
  \input{sections/related_work}

\section{Conclusion}
In this paper, we introduce MicroScopiQ, a novel co-design technique for post-training quantization of FMs. MicroScopiQ addresses outlier-aware quantization by: a) using higher-precision MX-FP for outliers and lower-precision INTs for inliers to maintain accuracy, and b) pruning to distribute additional outlier bits, ensuring memory alignment and hardware efficiency. The MicroScopiQ accelerator, featuring multi-precision PEs and a ReCoN NoC, efficiently handles distributed outliers. MicroScopiQ is the first PTQ technique to achieve SoTA compression for both LLMs and VLMs at just $2.36$-bits, offering $3\times$ better inference performance and $2\times$ lower energy consumption than existing accelerators.

\vspace{-2mm}
\begin{acks}
This work was supported in part by CoCoSys, one of the seven centers in JUMP 2.0, a Semiconductor Research Corporation (SRC) program sponsored by DARPA. The authors would like to thank the anonymous reviewers for their valuable feedback and suggestions, which helped improve the quality of this paper. The authors also thank Jiawei Hu for place-and-route, Jianming Tong for help with designing ReCoN, and Zishen Wan for functional accuracy simulation and constructive suggestions.  
\end{acks}

\bibliographystyle{ACM-Reference-Format}
\bibliography{refs}

\end{document}

%% file: includes/packages.tex

\usepackage{amsmath,amsfonts}
\usepackage{algorithmic}
\usepackage{graphicx}
\usepackage{textcomp}
\usepackage{xcolor}
\usepackage{multirow}
\usepackage{enumitem}
\usepackage{amsmath}
\usepackage{graphicx}
\usepackage{enumitem}
\usepackage{textcomp}
\usepackage{xcolor}
\usepackage{makecell}
\usepackage{multirow}
\usepackage{float}
\usepackage{fancyhdr}
\usepackage{colortbl}
\usepackage{arydshln}
\usepackage{graphicx}
\usepackage{textcomp}
\usepackage{xcolor,soul}
\usepackage{hyperref}
\usepackage{wrapfig}
\usepackage{algorithm2e}
\RestyleAlgo{ruled} 
\usepackage{pifont}
\usepackage{amsmath}
\usepackage{graphicx}
\usepackage{textcomp}
\usepackage{xcolor}
\usepackage{makecell}
\usepackage{multirow}
\usepackage{tikz}
\usepackage{float}
\usepackage{graphicx}
\usepackage{booktabs}
\usepackage{cleveref}
\crefformat{section}{\S#2#1#3} 
\crefformat{subsection}{\S#2#1#3}
\crefformat{subsubsection}{\S#2#1#3}

\newcommand{\tick}{\includegraphics[height=1.5ex]{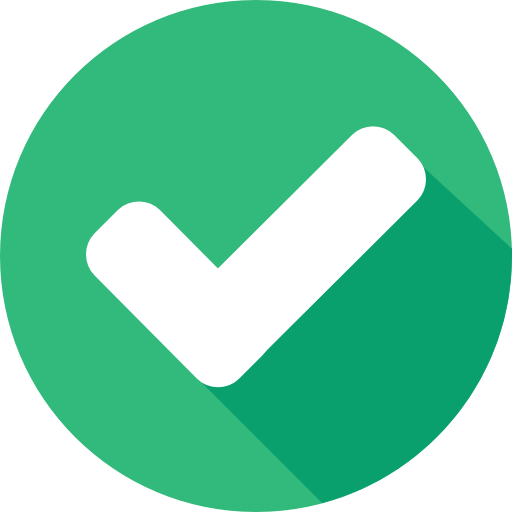}}
\newcommand{\cross}{\includegraphics[height=1.5ex]{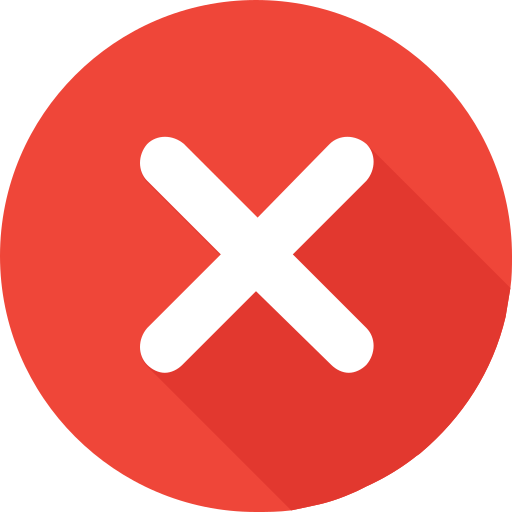}}
\newcommand{\exclamation}{\includegraphics[height=1.5ex]{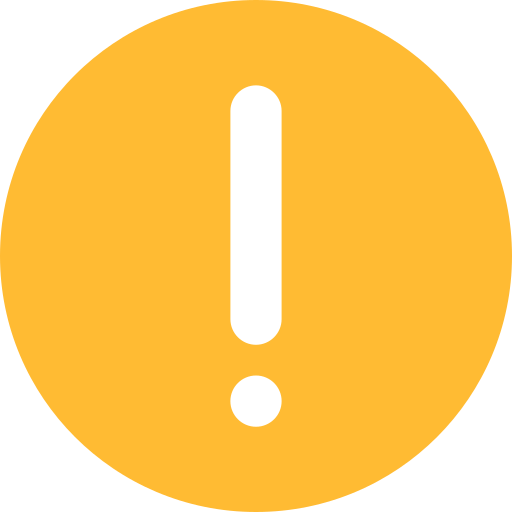}}
\newcommand*\circled[1]{\tikz[baseline=(char.base)]{
            \node[shape=circle,fill,inner sep=0pt] (char) {\textcolor{white}{#1}};}}

%% file: sections/introduction.tex
\section{Introduction}
\label{sec:introduction}
Recent advancements in AI \cite{goodfellow2014generative, li2022efficient, touvron2023llama, seal2025chen} have been propelled by a class of models called foundational models (FMs), which encompass large language models (LLMs) and vision-language models (VLMs). FMs leverage billion-scale parameters for improved learning \cite{yang2024harnessing, zhang2024scaling} but impose substantial demands on memory, energy, and compute resources. Recent research has focused on various model compression techniques such as pruning \cite{frantar2023sparsegpt, ashkboos2024slicegpt, yin2023outlier} and quantization \cite{lin2024awq, shao2023omniquant, ramachandran2024algorithm} to reduce memory and computational overhead, enabling efficient FM inference on resource-constrained devices.

\textit{Model pruning} \cite{frantar2023sparsegpt} reduces memory footprint by removing ineffectual model parameters, such as individual weights (unstructured) or blocks of weights (structured), and storing sparse tensors in a compressed format \cite{jeong2023vegeta}. However, pruning of FMs may be infeasible due to, significant accuracy drops even at low pruning ratios \cite{kuzmin2024pruning, yinjunk} and potential demand for compute and memory-intensive fine-tuning to regain accuracy.
\textit{Model quantization}, on the other hand, reduces the size of a target model by representing weights and/or activations at low precision \cite{ramachandran2024algorithm, kang2024gear, ramachandran2024clamp, kundu2025lvlm}. Recent works on quantization \cite{dettmers2022llm, guo2023olive, ramachandran2025ouromamba} have identified that quantizing LLMs is considerably more challenging than quantizing traditional DNNs \cite{li2021brecq, tambe2020algorithm} due to the emergence of large magnitude features known as \emph{outliers} \cite{zadeh2020gobo}. These outliers significantly impact model accuracy and require specialized handling \cite{guo2023olive} compared to inliers.

\input{tables/intro_comparison}

To address the issue of outliers in FMs, recent algorithm/ architecture co-design techniques \cite{shao2023omniquant, lin2024awq, guo2023olive} have proposed different types of outlier-aware quantization. These techniques can be broadly categorized based on their outlier handling approach: \circled{A} Maintaining outliers at higher precision compared to the inliers, or \circled{B} Quantizing outliers at the same precision as inliers while using different data formats or scaling factors for outliers.

Techniques in group \circled{A}, such as OWQ \cite{lee2024owq}, SpQR \cite{dettmers2023spqr}, SDQ \cite{jeong2024sdq} (algorithm) and GOBO \cite{zadeh2020gobo}, OLAccel \cite{park2018energy}(architecture co-design) exhibit low accuracy degradation. This is because, they typically store outliers at high precision separated from lower precision inliers. However, these techniques result in, (a) low compression factor with high \textit{effective bit-width} (EBW\footnote{The average number of bits used to represent each quantized parameter of a model.})  and (b) inefficient hardware and unaligned memory access.

On the other hand, techniques in group \circled{B}, such as AWQ \cite{lin2024awq} (algorithm) and OliVe \cite{guo2023olive} (architecture co-design) quantize outliers at the same precision as inliers following different strategies. AWQ tries to identify a separate outlier-specific scale factor via channel-wise scaling. OliVe uses the ``flint'' data format \cite{guo2022ant} for inliers and ``abfloat'' \cite{guo2023olive} for outliers, both at 4-bit precision. These techniques mitigate the unaligned memory access while providing high compression. However, they suffer from significant accuracy degradation, particularly at ultra-low bit widths. This may be attributed to the reduced representational range available to outliers at ultra low-precision. Additionally, these methods \cite{guo2023olive} rely on a specific kind of locality of presence for outliers, that might not be true for all FMs, as we shall demonstrate in this work (\cref{sec:inefficiency_background}).


Based on the shortcomings of existing solutions discussed above, we identify that assignment of higher bit-width for outliers is essential for good accuracy while for aligned memory and hardware efficiency a consistent \textit{bit-budget} and data type per tensor element is desired. Here, by consistency we mean that on average each scalar within a tensor should be represented by a fixed bit-width of a particular data-type. However, these demands are conflicting.

\noindent
\textbf{Contributions.} To provide a unified solution, we investigate on a fundamental question: 
\begin{mybox}
\noindent
\textbf{Question:} \textit{Can pruning be effectively leveraged to complement outlier-aware quantization in achieving high accuracy while maintaining hardware efficiency?}
\end{mybox}


Towards achieving this, we present a novel co-design technique for the post-training quantization (PTQ) of FMs, namely \textbf{MicroScopiQ}. Our approach effectively leverages pruning with outlier aware quantization to achieve both memory alignment and improved accuracy. To effectively perform this for a layer, we quantize outliers at twice the precision of inliers and prune the least important weights based on the Hessian information. We then redistribute the additional bits of the outlier weights in these pruned locations. This ensures memory alignment while allowing outlier weights to have higher precision. Additionally, to reduce error we use the recently proposed MicroScaling (MX) FP data format \cite{rouhani2023microscaling, darvish2023shared} to quantize outlier weights as opposed to MX-INT inlier quantization. While prior work such as SDQ \cite{jeong2024sdq} also combined pruning and quantization, it contrasts with our approach in its limited outlier flexibility, lower compression factor, and unaligned memory access.

To efficiently support outliers in a \textbf{different} format with \textbf{different} bits at \textbf{different} locations in hardware, we present an intelligent NoC architecture called, \underline{Re}distribution and \underline{Co}ordination \underline{N}oC (ReCoN). It offers minimal overhead and high throughput outlier processing and reorganization. We then present an accelerator that leverages ReCoN with a simple, homogeneous INT-PE array. Additionally, we extend the accelerator to be generic enough to support multiple bit-precision (2/4-bit operations). As summarized in \autoref{tab:intro-comparison}, MicroScopiQ blends the advantages of group \circled{A} and \circled{B} techniques while mitigating their specific drawbacks.

Our key contributions can be summarized as follows:

\begin{itemize}[align=right,itemindent=2em,labelsep=2pt,labelwidth=1em,leftmargin=0pt,nosep]
   \item We present MicroScopiQ, a PTQ framework to efficiently integrate pruning with outlier-aware quantization (\cref{sec:algo}).

  \item To effectively deploy MicroScopiQ in a systolic array architecture we present a novel architecture supporting multi-precision, homogeneous PEs  with a low-overhead NoC architecture (\cref{sec:hardware}).

  \item To our best knowledge, MicroScopiQ is the first co-design technique, to push the limits of PTQ compression for \textbf{both} LLMs and VLMs with an EBW of $\sim \hspace{-4pt} \textbf{2.36}$\textbf{-bits} for weights; achieving SoTA quantized model accuracy across different weight/weight-activation quantization settings. Moreover, it demonstrates up to $\textbf{3}\times$ \textbf{improvement in performance-per-unit-area} ($TOPS/mm^2$) and up to $\textbf{35}\%$ \textbf{energy reduction} compared to existing architectures (\cref{sec:results}).
\end{itemize}

%% file: tables/intro_comparison.tex
\begingroup	
\begin{table}[t]\centering
 \caption{ MicroScopiQ vs. prior outlier-aware quantization techniques, categorized into two groups, A~\cite{zadeh2020gobo, park2018energy}, B~\cite{guo2023olive}.}
 \vspace{-5pt}
\resizebox{\linewidth}{!}
{%
\begin{tabular}{l|lll}
\Xhline{2\arrayrulewidth}
\rowcolor[HTML]{E0E0E0}
 & \multicolumn{3}{c}{\textbf{Methods}} \\
\cline{2-4}
\rowcolor[HTML]{E0E0E0}
\raisebox{1.2ex}[1.2ex]{\textbf{Categories}}& Group \circled{A} & Group \circled{B} & \textbf{MicroScopiQ} \\
\Xhline{1\arrayrulewidth}
Accuracy & \tick \hspace{0.07em} High & \cross \hspace{0.07em} Low & \tick\hspace{0.07em} \textbf{High} \\
\Xhline{1\arrayrulewidth}
Effective bit-width & \cross\hspace{0.07em} High (18.17b) & \tick\hspace{0.07em} Low (2b) & \tick\hspace{0.07em} \textbf{Low} \textbf{(2.36b)} \\
\Xhline{1\arrayrulewidth}
Flexibility & \cross\hspace{0.07em} No & \cross\hspace{0.07em} No & \tick\hspace{0.07em} \textbf{Yes} \\
\Xhline{1\arrayrulewidth}
Aligned memory & \cross\hspace{0.07em} Unaligned & \tick\hspace{0.07em} Aligned & \tick\hspace{0.07em} \textbf{Aligned} \\
\Xhline{1\arrayrulewidth}
PE design & \cross\hspace{0.07em} Complex & \cross\hspace{0.07em} Complex & \tick\hspace{0.07em} \textbf{Simple} \\
\Xhline{1\arrayrulewidth}
HW overhead & \cross\hspace{0.07em} High & \exclamation\hspace{0.07em} Moderate & \tick\hspace{0.07em} \textbf{Low}  \\
 \Xhline{2\arrayrulewidth}
\end{tabular}
}
\vspace{-6mm}
\label{tab:intro-comparison}
\end{table}
\endgroup

%% file: sections/background_motivation.tex
\section{Background}
\subsection{Model Quantization}
\label{sec:quantization}
A typical quantization \cite{ramachandran2024clamp, li2021brecq, ramachandran2025ouromamba} process involves two steps: establishing quantization parameters given the quantization data format $(\tau)$ and bit-width $(b)$, and mapping the high-precision tensor to the quantized representation. For a typical symmetric quantization \cite{zhao2024atom} (zero-point is 0) of a tensor $X$, the  \emph{scale factor} ($s$) is given by,
\vspace{-1mm}
\begin{equation}
\label{equation:scale_factor}
    s = \frac{\emph{max}(X)}{\emph{max}_{\tau}^b}
\end{equation}
\vspace{-2mm}

\noindent
$\emph{max}_{\tau}^b$ is the maximum representable value of a data format \cite{ramachandran2024clamp}. For $b$-bit INT quantization, $\emph{max}_{INT}^b = 2^{b-1}-1$. After determining the quantization parameters, the quantized tensor is given by \cite{ramachandran2024clamp},

\begin{equation}
\label{equation:scaling}
   Q(X, s, b) = \emph{clip}(\Bigl\lfloor{\frac{X}{s}}\Bigr\rceil, \emph{min}_{\tau}^b, \emph{max}_{\tau}^b) 
\end{equation}
\vspace{-3mm}

\noindent
In model quantization, the quantization parameters can be shared at different granularity for different accuracy-overhead trade-offs. In increasing order of overheads, we have \textbf{per-tensor} quantization, wherein the scale factor  is shared among all tensor elements. In \textbf{per-channel} quantization, the scale factor is shared per row/column of a tensor. Finally, in \textbf{group} quantization, the parameters are shared at a finer granularity between groups of $k$ (64, 128 etc.) elements in a row or column. These groups are formed by dividing channels into multiple non-overlapping contiguous blocks. \emph{In this paper, we adopt MX-INT and MX-FP quantization for inliers and outliers, respectively.}

\begin{figure}[t]
    \centering\includegraphics[width=0.95\columnwidth, keepaspectratio]{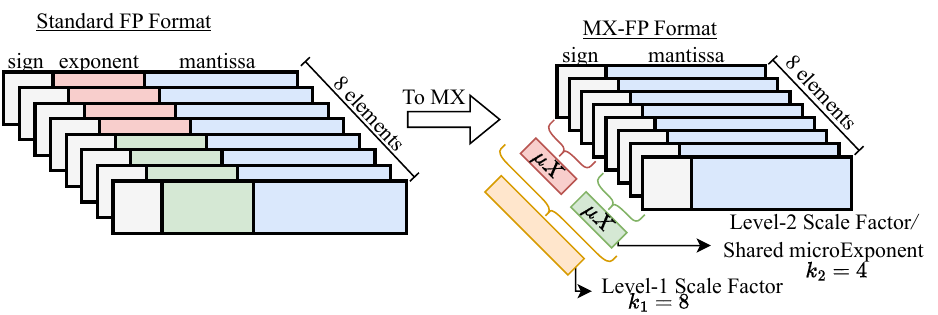}
        \vspace{-4mm}
        \caption{Depiction of MX-FP data format with level-1 scale factor and level-2 microExponent  ($\mu X$), with $k_1$ and $k_2$ the group sizes over which these two factors are shared.}
        \label{fig:mx_format}
        \vspace{-6.5mm}
\end{figure}
\subsection{Microscaling Data Format}
\label{sec:background_mx}
The MX data format proposed by prior works \cite{rouhani2023microscaling, darvish2023shared}, is standardized by the Open Compute Project~\cite{ocp2024} with support from Microsoft, Intel, NVIDIA and others. As shown in \autoref{fig:mx_format}, MX is a variant of block data representation (BDR) \cite{darvish2023shared} that defines a format to represent a group of values collectively using shared scale factors. It leverages multi-level, power-of-two scaling, at fine- (level-1, $k_1$) and ultra-fine (level-2, $k_2$) granularity \cite{dai2021vs, drumond2018training}. The MX data format is characterized by four components: i) scale factors (level-1, 2), ii)  data type ($\tau$), iii) bit-width ($b$) and iv) group sizes ($k_1, k_2$). In this paper we denote an MX-FP format as MX-FP-$b$$_{k_1,k_2}$. In this work, we adopt the version of the MX-FP data format proposed in \cite{rouhani2023microscaling}, employing multi-level scaling. The level-1 scale factor for MX-FP is computed following \autoref{equation:scale_factor}. Conversely, for level-2 scale factor, we \emph{identify} that MX-FP leverages the sharing of exponent field of FP values \cite{wang2018training} (referred as $\mu X$ in \autoref{fig:mx_format}). \emph{We show in \cref{sec:inlier_outlier} that by taking advantage of this insight i.e., the concept of shared $\mu X$, we are able to represent FP-outliers in INT format, thereby, enabling the design of simple, homogeneous INT-based PEs.} For inliers, we employ MX-INT-$b$$_{k_1}$ with a single level of scale factor following \cite{ocp2024}. This is because, INT format does not possess an exponent field, thereby, a level-2 scale factor similar to MX-FP is not applicable. For simplified understanding, MX-INT-$b$$_{k_1}$ inlier quantization can be viewed as analogous to INT group quantization utilizing an E8M0 scale factor.
\vspace{-3mm}
\begin{figure}[t]
    \centering

    \includegraphics[width=\columnwidth, keepaspectratio]{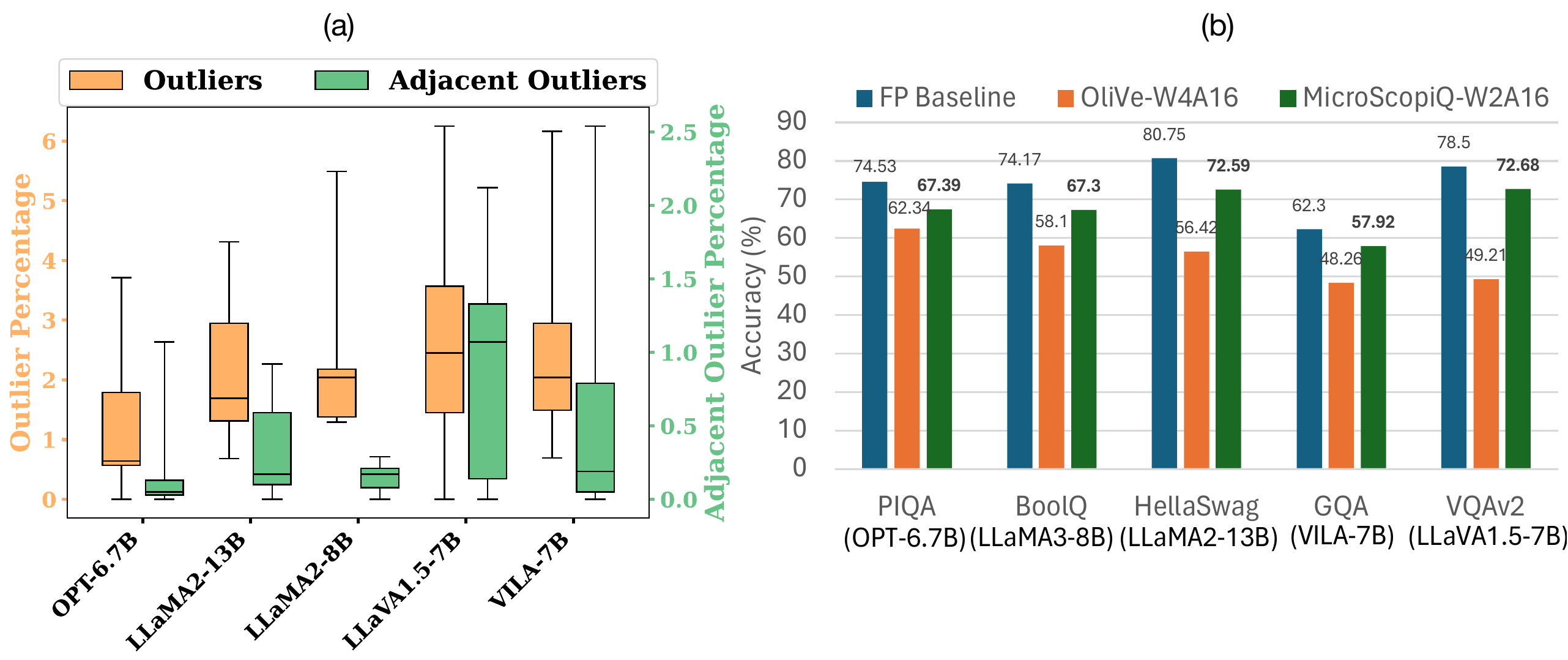}
        \vspace{-8mm}
        \caption{(a) Layer-wise distribution of outliers and adjacent outliers as a percentage of total number of weights, (b) Quantization accuracy comparison between OliVe-W4A16 and MicroScopiQ-W2A16 on various benchmarks.}
        \label{fig:outlier_motivation}
        \vspace{-7mm}
\end{figure}

\section{Motivation}
\subsection{Limitations of existing techniques}
\label{sec:inefficiency_background}
In \autoref{tab:intro-comparison}, we compare candidate proposals from group \circled{A}: GOBO \cite{zadeh2020gobo} and group \circled{B}: OliVe \cite{guo2023olive} across various metrics. GOBO is able to achieve high accuracy by retaining outliers at full-precision. It stores outliers separately from low-precision inliers by using sparse representations with the associated outlier indices (see \autoref{fig:algo_fig}(b)). By retaining outliers at full-precision, GOBO results in high EBW. Moreover, the compressed sparse storage and multiple precisions results in unaligned and random memory accesses \cite{guo2023olive}, significantly impacting inference latency. Furthermore, GOBO's outlier handling is hardware inefficient, requiring complex PEs. Similarly, a recent work \cite{jeong2024sdq} proposed to decompose a vector of weights into two separate inlier and
outlier vectors each quantized in different precisions with outliers at a higher precision.

OliVe \cite{guo2023olive} proposes a scheme to ensure aligned memory access by quantizing inliers and outliers at the same precision (low EBW), but using different data formats. To enable differentiation between the inlier and outlier formats, it prunes the value adjacent to the outlier for use as an identifier (see \autoref{fig:algo_fig}(c)). 
However, OliVe results in significant accuracy degradation, especially at low precision (see \autoref{fig:outlier_motivation}(b)), due to: 1) sacrificing a number encoding from inliers for exclusive use as an identifier, reducing the number of representable values in the quantized range, and 2) the rigid assumption of outlier locality--that outliers are never adjacent to each other and only inliers are almost always adjacent to outliers (see \cref{sec:motivation_1}), leading to unintended outlier pruning. Furthermore, OliVe requires a fairly complex PE design incurring significant encoding/decoding overheads to convert the different formats into a unified processing format (exponent-integer pair). \emph{In this paper, we show that despite quantizing outliers at higher-precision and in a different format, we ensure aligned memory access, simple PE design and minimal hardware overhead.}

\subsection{Adjacent Outliers Matter}
\label{sec:motivation_1}

Similar to prior works \cite{guo2023olive, park2018energy}, we leverage the $3\sigma$ rule \cite{pukelsheim1994three} to categorize weights as outliers. We visually demonstrate the distribution of outliers and adjacent outliers \footnote{We define adjacent outliers as two contiguous outliers along the dot-product dimension (see row 2 of the LLM weight matrix in the center of \autoref{fig:algo_fig}).} as a percentage of the total number of weights in a layer across different FMs in \autoref{fig:outlier_motivation}(a). As the \textcolor{orange}{orange} box-plot shows, outliers depict a maximum  percentage of $\sim \hspace{-3pt} 5.1\%$. Outliers are prevalent in FMs, and preserving their values is crucial for maintaining quantized model accuracy. \textbf{Importantly}, from the \textcolor[HTML]{006400}{green} box-plots, we observe that modern FMs on average possess $>0.5\%$ adjacent outliers per layer, with some FM layers showing peaks of $>2\%$. This is in stark contrast to the models evaluated by OliVe, such as BERT \cite{devlin2018bert} and OPT \cite{zhang2022opt} which have $<0.04\%$ adjacent outliers (two orders of magnitude lower than FMs like LLaMA3 and LlaVa). This indicates that while pruning values adjacent to outliers could have been ideal for models like BERT \cite{devlin2018bert}, it is sub-optimal for most modern FMs as it removes crucial outlier values, leading to higher accuracy degradation. This is evident from \autoref{fig:outlier_motivation}(b) where OliVe has significant accuracy degradation at 4-bit quantization due to its assumption on outlier locality. \emph{Unlike OliVe, MicroScopiQ does not naively prune adjacent values; instead it leverages the Hessian information \cite{frantar2022gptq} to identify the least important values to prune, ensuring outlier preservation. This directly translates to high quantized model accuracy and MicroScopiQ at 2-bit consistently outperforms OliVe across different FMs.}

\vspace{-2mm}
\subsection{Outlier Precision and Data Format}
\label{sec:motivation_2}
The ability of group \circled{A} techniques like GOBO \cite{zadeh2020gobo} to achieve high quantized model accuracy even at extreme quantization levels of inliers ($<4$-bits) is due to retaining outliers at higher precision. This is particularly crucial at ultra-low bit width quantization because, if inliers and outliers are to be quantized to the same precision, there will be higher outlier quantization error due to the reduced representational range. We demonstrate this effect on the MicroScopiQ quantized FM accuracy in \autoref{tab:ablation_2} wherein the quantized FM has poor performance when inliers and outliers are at 2-bits compared to outliers at 4-bits. Furthermore, evidence from recent work \cite{wu2023zeroquant} demonstrates that FP-based formats for LLMs results in superior quantization performance compared to INTs. To validate this, we compare MX-INT v/s MX-FP inlier and outlier quantization in \autoref{tab:ablation_2}. Evidently using MX-FP instead of MX-INT for outliers results in better performance. This is due to the higher dynamic range of FPs, which is particularly beneficial at extreme quantization levels. \emph{In this work, we quantize outliers at a higher precision (2$\times$) compared to inliers, using MX-FP for outliers and MX-INT for inliers.}

%% file: sections/algo_methodology.tex
\vspace{-1mm}
\section{MicroScopiQ Quantization Methodology}
\label{sec:algo}

\begin{figure}[t]
    \centering
    \includegraphics[width=\linewidth, keepaspectratio]{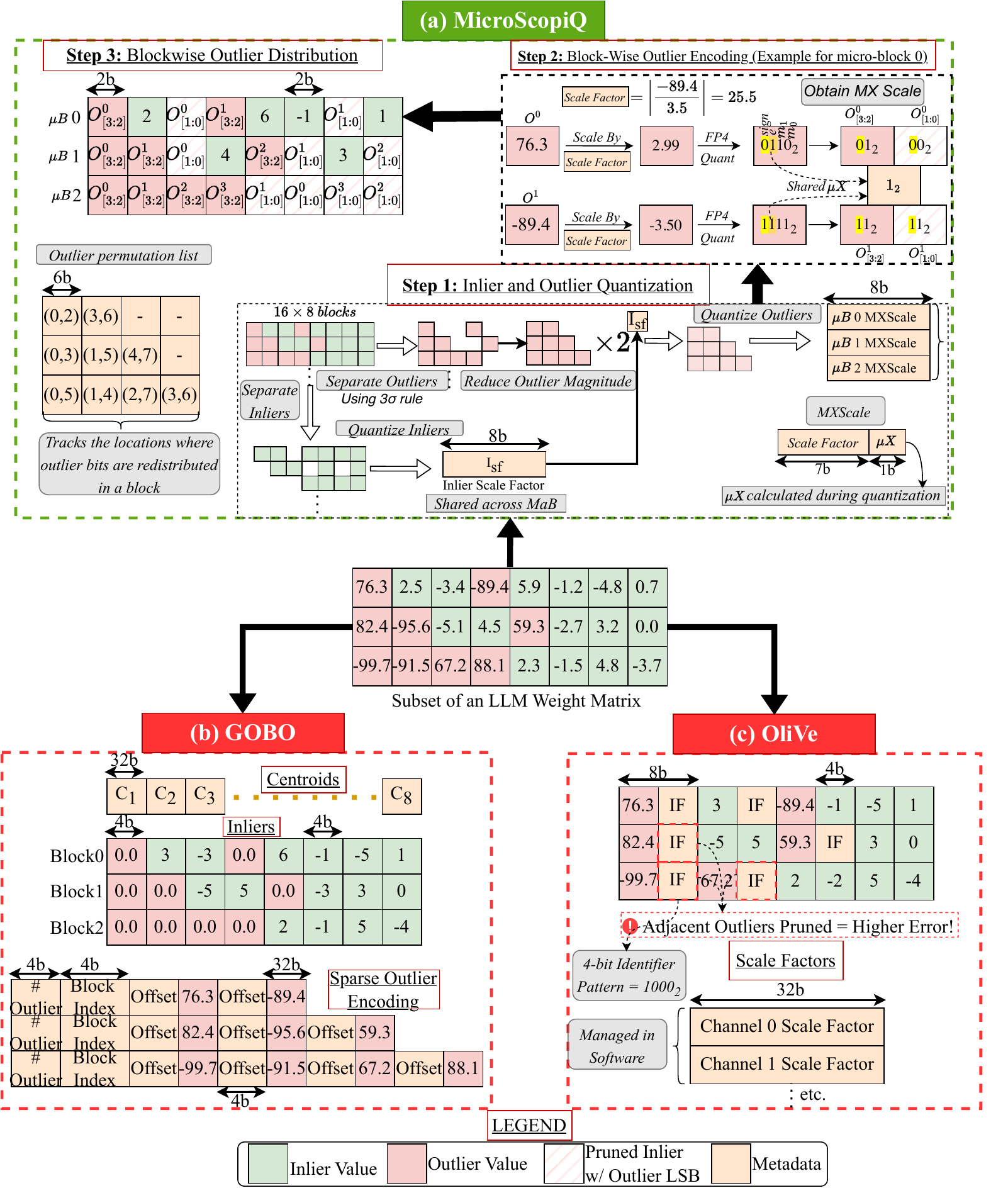}
        \vspace{-6mm}
        \caption{(a) Overview of the proposed MicroScopiQ quantization framework depicting methodology of inlier and outlier quantization and redistribution of outlier bits for a sample LLM weight matrix. Comparison against prior quantization frameworks (b) GOBO, and (c) OliVe.} 
        \label{fig:algo_fig}
        \vspace{-5mm}
\end{figure}

We present an overview of MicroScopiQ quantization in \autoref{fig:algo_fig}(a) and detail it in Algorithm \ref{algo:microscopiq}. MicroScopiQ supports various group size granularities and any inlier and outlier ($2\times$ inliers) data precision. For simplicity, we explain with inlier and outlier precision of 2/4- and 4/8-bit and group sizes of 128 for inliers and 8 for outliers.

\subsection{Preliminaries}
\label{sec:preliminaries}
The MicroScopiQ quantization framework models the layer-wise post-training quantization of FMs by partitioning each layer into multiple rows and quantizing each row at a time. Concretely, for a given input calibration dataset $\mathbb{X}$, at layer $l$ the objective is to find a quantized set of weights $\mathbb{Q} \in \mathcal{R}^{d_{row} \times d_{col}}$ that minimizes the sum of squared errors over all rows of the layer compared to the full-precision weights $\mathbb{W} \in \mathcal{R}^{d_{row} \times d_{col}}$. This can be formulated as, 

\vspace{-3mm}
\begin{equation}
    \textbf{argmin}_{\mathbb{Q}} \sum_{i=1}^{d_{row}} ||\mathbb{W}_{i,:}\mathbb{X} - \mathbb{Q}_{i,:}\mathbb{X} ||_2^2
    \label{equation:quadratic}
\end{equation}
\vspace{-2mm}

We evaluate the second order derivative of \autoref{equation:quadratic}, namely the Hessian \cite{hassibi1993optimal, lecun1989optimal} through Taylor series expansion \cite{hassibi1993optimal}. Note, the Hessian is the same for all rows due to its dependence only on input data and is given as, $\mathbb{H} = 2\mathbb{X}\mathbb{X}^T$. Its inverse is, $\mathbb{H}^{-1} = (2\mathbb{X}\mathbb{X}^T + \lambda\mathbb{I})^{-1}$. We leverage the Hessian information to identify weights with small ``saliencies" for pruning (L17 in Algo. \ref{algo:microscopiq}). The number of inliers to prune is determined by the number of outliers, as detailed later.  

To improve quantization performance, we take inspiration from \cite{frantar2022gptq} and adjust the weights of unquantized rows to minimize the net error while quantizing a particular row of weights. The associated equation (L31 of Algo. \ref{algo:microscopiq}) for weight update using the Hessian is derived by solving the Lagrangian of \autoref{equation:quadratic}. However, updating all remaining rows each time a row is quantized incurs significant compute-overhead, making this intractable for billion-scale FMs. Therefore, as pointed out in \cite{frantar2022gptq}, we partition the rows into non-overlapping contiguous row-blocks (rB) of size $128$ rows and localize the updates of unquantized rows within each rB. We only update the rows outside a rB (L36 in Algo. \ref{algo:microscopiq}) once all the elements of the current block are quantized. This minimizes the number of individual updates by grouping updates together per rB and producing an order of magnitude speedup.

\input{algo/microscopiq_algo}

\vspace{-2mm}
\subsection{Inlier and Outlier Weight Quantization}
\label{sec:inlier_outlier}

In \textcolor{blue}{\textbf{Step 1}} (\autoref{fig:algo_fig}(a), Algorithm \ref{algo:microscopiq}), each row to be quantized is first divided into multiple non overlapping contiguous \textbf{macro-blocks (MaBs)} of size $B_M = 128$. All inliers are quantized within a MaB and share the scale factor. Each MaB is then subdivided into multiple non-overlapping contiguous \textbf{micro-blocks ($\mu$Bs)} of size $B_{\mu} = 8$ with sixteen $\mu$Bs forming a MaB. The outliers present in each $\mu$B shares same scale(s). As depicted in \autoref{fig:algo_fig}(a), \textcolor{blue}{\textbf{Step 1}}, the quantization process begins by first identifying inliers and outliers in each MaB by using the $3\sigma$ rule. A shared 8-bit power-of-two scale factor ($2^{I_{sf}}$), following \autoref{equation:scale_factor} is calculated for all inliers in a MaB and the inliers are quantized to 2-bit or 4-bit, resulting in MX-INT-(2/4)$_{128}$ quantization. Interestingly, \textbf{we observe that the inlier scale factor in each MaB is always a \emph{negative} power of two} for all FMs under consideration. 
We leverage this observation to \textbf{reduce outlier magnitude}, by multiplying all outlier values in a MaB with the inlier scale factor ($2^{I_{sf}}$) (this can also be perceived as division by $2^{-I_{sf}}$, for conformity with \autoref{equation:scaling}). This pre-processing helps make outlier quantization easier, by pre-reducing its dynamic range before the actual outlier quantization. 

Unlike inliers, outliers are quantized per $\mu$B, to reduce quantization error due to shared scaling over a larger group size (see \cref{sec:results}). After identifying outliers present in a $\mu$B, we compute a shared 8-bit MXScale that is calculated by concatenating the level-1 power-of-two scale factor ($2^{O^{l_1}_{sf}}$) and level-2 microExponent ($\mu X$). The level-1 scale factor is calculated by following \autoref{equation:scale_factor} to obtain 7 or 5-bit MSBs of MXScale depending on size of $\mu X$ being 1 or 3-bit of the LSBs--corresponds to exponent size of the FP format (depicted in \textcolor{blue}{\textbf{Step 2}} in \autoref{fig:algo_fig}(a)). The outliers in a $\mu$B are scaled by ($2^{O^{l_1}_{sf}}$), following \autoref{equation:scaling} and then quantized to either \emph{e1m2}/\emph{e3m4} FP-format \cite{rouhani2023microscaling} for $b$ = 4 or 8-bit, respectively. Post quantization of outliers, the level-two scale factor or the $\mu X$ is obtained by extracting the common exponent among all outliers in a $\mu$B. This process results in a MX-FP-$b$$_{8,8}$ outlier quantization. The final outlier scale factor is $2^{O_{sf}}$ where, $O_{sf}$ is expressed as ${O_{sf}} = {O^{l_1}_{sf} + \mu X - I_{sf}}$. The term $I_{sf}$ in the final outlier scale factor accounts for multiplication by $2^{I_{sf}}$ (or division by the inverse) during outlier pre-processing. 
\vspace{-2mm}

\subsection{Outlier Value Encoding through N:M Structured Pruning}
\label{sec:outlier_pruning}
To formalize \textcolor{blue}{\textbf{Step 2}}, let us assume $n$ outliers are present in a $\mu$B quantized to MX-FP-$4$$_{8,8}$. After sharing 1-bit $\mu X$ across the $n$ outliers, each outlier has 1 sign ($s$) and 2 mantissa bits ($m_1m_0$). The inliers are 2-bit 2's complement MX-INT and has 1-bit sign and 1-bit magnitude. Since the outlier bits have two mantissa bits with one sign, to ensure symmetric outlier distribution, we duplicate their sign bit and assign each sign to a mantissa creating and partitioning into two halves \texttt{Upper,Lower} of size 2-bit each mimicking inlier MX-INT structure i.e., $\{sm_1, sm_0\}$ (see \autoref{fig:algo_fig}(a), \textcolor{blue}{\textbf{Step 2}}). While we demonstrate this for a specific example, the process can be generalized to other inlier and outlier formats bit-widths wherein, the remaining outlier bits need to be reorganized into \texttt{Upper} and \texttt{Lower} distributable halves each of size $bb$ (\textit{per element bit-budget}), with $bb$ representing the operational bit-width of each PE.

In \textcolor{blue}{\textbf{Step 3}}, to ensure a fixed $bb$ and data-type of a layer for aligned memory access and simple PE design, we distribute the outlier LSBs (\texttt{Lower} half) to the least important inlier locations that are pruned within a $\mu$B. For $n$ outliers in a $\mu$B we identify $n$ least important inlier locations to prune via the Hessian information (see Algorithm \ref{algo:microscopiq}), forming a ($B_{\mu}$-$n$):$B_{\mu}$ structured pruning pattern \footnote{Following the structure of a standard N:M pattern \cite{jeong2023vegeta} where N = ($B_{\mu}$-$n$) and M = $B_{\mu}$} \cite{jeong2023vegeta, sun2021dominosearch} i.e., ($B_{\mu}$-$n$) non-zero values exist for every $B_{\mu}$ after pruning. To keep track of the corresponding halves of each of the outliers in a $\mu$B, we maintain a per-$\mu$B permutation list (Note: if there are no outliers in a $\mu$B, a permutation list is not stored). The permutation list for each $\mu$B that has outliers is made up of $\frac{B_{\mu}}{2}$ elements (maximum number of outliers supported per $\mu B$) with each element storing the locations of the \texttt{Upper} and \texttt{Lower} halves of outliers in a 6-bit format $\{\texttt{Upper}_{loc}, \texttt{Lower}_{loc} \}$ (see \autoref{fig:algo_fig}(a) \textcolor{blue}{\textbf{Step 3}}). Note that, in case of outlier count $ > \frac{B_{\mu}}{2}$, the mB size should be chosen to be higher to prevent pruning of outliers and causing higher accuracy degradation. While such a situation does not arise in any of the models that we evaluate with $B_{\mu}=8$, this feature only serves as a demonstration of the flexibility of MicroScopiQ for future models. 

\begin{figure}[t]
\vspace{-0.5mm} 
    \centering
    \includegraphics[width=\columnwidth, keepaspectratio]{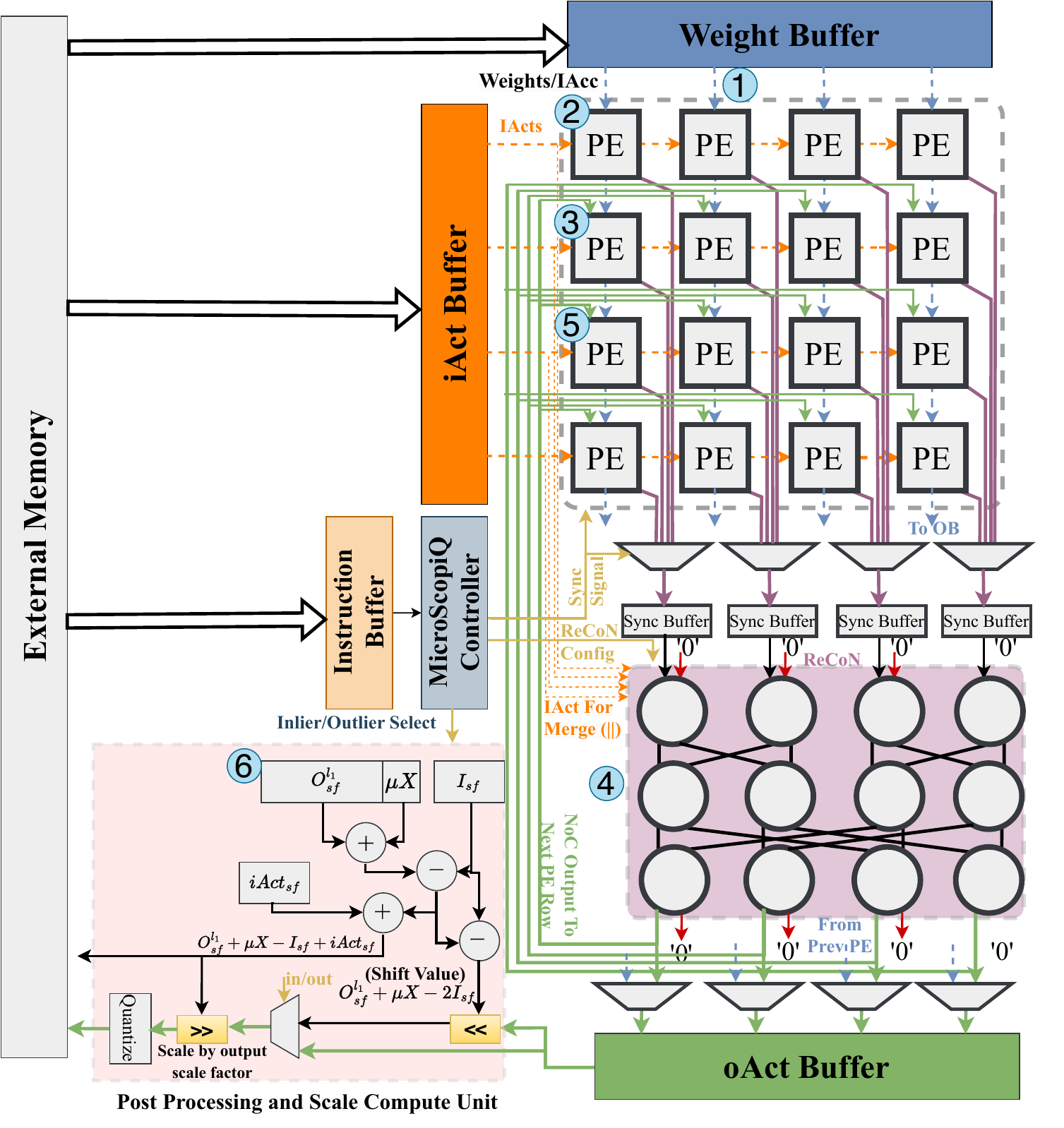}
        \vspace{-8mm}
        \caption{Integration of MicroScopiQ into a WS systolic array.}
        \label{fig:hw_architecture}
\end{figure}

\begin{figure}[t]
    \centering
    \includegraphics[width=\columnwidth, keepaspectratio]{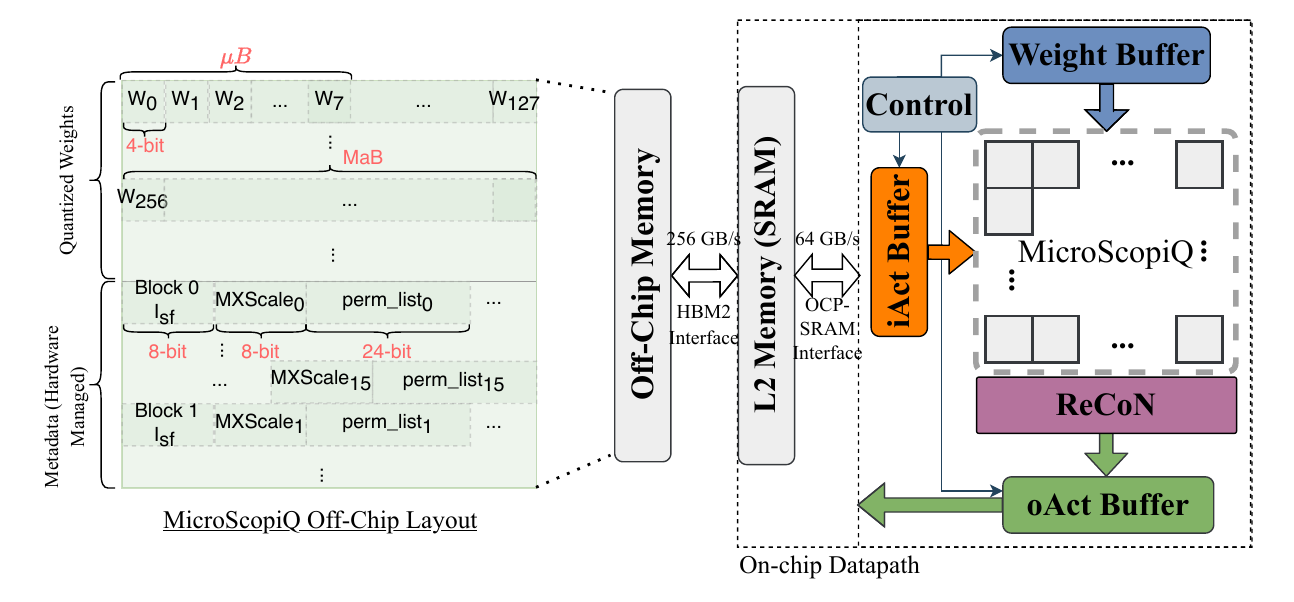}
        \vspace{-10mm}
        \caption{MicroScopiQ memory organization.}
        \label{fig:memory_layout}
        \vspace{-1mm}
\end{figure}

\subsection{Effective Bit Width (EBW) Calculation}
Following \cite{frantar2022gptq, zhao2024atom} we report the EBW as the average number of bits required for storing each element including the metadata. The EBW of MicroScopiQ quantized FM varies dynamically across models and is dependent on the outlier percentage and choice of $\mu$B size. For INT-$2_{B_{M}}$ inlier and MX-FP-4$_{B_{\mu}, B_{\mu}}$ outlier quantization, if there are no outliers in a $\mu$B the EBW of that $\mu$B is $\texttt{EBW}_I = 2$, i.e., $bb$, whereas if there are outliers present in a mB the EBW is $\texttt{EBW}_O = (perm^{bits} + 2B_{\mu} + O_{sf}^{bits})/{B_{\mu}}$, which translates to 6-bits for $B_\mu$=8, MXScale of 8-bits and permutation list size of 24-bits (see \cref{sec:outlier_pruning}). Since the inlier scale factor is shared across a larger group size $B_M$, its contribution to the EBW is negligible and hence ignored. As a rule of thumb, if there are no outlier present in a $\mu$B the EBW of that $\mu$B is equal $bb$ else the EBW includes the contribution of all outlier-specific metadata. Additionally, the presence or absence of outlier metadata is delineated by a 1-bit identifier per $\mu$B, wherein a $1$ indicates presence of outlier metadata. This identifier contributes negligible overhead to the EBW (0.05-0.09 bits) similar to the inlier scale factor and is hence ignored in the EBW calculation. For a FM with $l$ layers and each layer having $m$ $\mu$Bs and $x\%$ of these $\mu$Bs consist of outliers, the EBW of the FM is,

\begin{equation}
\texttt{EBW}_{FM} = \frac{\sum_{i=1}^l {(x \cdot m \cdot \texttt{EBW}_O + (100-x) \cdot m \cdot \texttt{EBW}_I)}/{m}}{l}
\end{equation}

%% file: algo/microscopiq_algo.tex
\newcommand{\ccomment}[2][black]{\textcolor{#1}{\emph{\# \textbf{#2}}}}
\SetAlFnt{\scriptsize}
\SetAlgoNlRelativeSize{-1} 
\LinesNumbered 
\SetNlSty{}{\scriptsize}{}
\setlength{\textfloatsep}{0pt}
\begin{algorithm}[t]
    \SetKwInOut{Input}{Input}
    \SetKwInOut{Output}{Output}
    \caption{MicroScopiQ Quantization Framework 
    }
    \label{algo:microscopiq} 
    \Input{$\mathbb{W} \in \mathcal{R}^{d_{row} \times d_{col}}$, calibration data $\mathbb{X}$, $\mathbb{H}^{-1} = (2\mathbb{X}\mathbb{X}^T + \lambda\mathbb{I})^{-1}$, row block (\textit{rB}), macro-block (\textit{MaB}) $B_{M}$, and micro-block (\textit{$\mu$B}) $B_{\mu}$}

    \Output{Quantized weight $\mathbb{Q} \in \mathcal{R}^{d_{row} \times d_{col}}$, perm (Permutation list)}
\ccomment[blue]{Iterate over row blocks}\

    \For{$i = 0, rB, 2rB, \cdots d_{col}-rB$}{
        \For{$j = i, i+1,\cdots, i+rB-1$}{
        \ccomment[blue]{Step 1.0: Divide each row into non-overlapping Macro-Blocks}\

            \For{$\mathbb{W}_{j,MaB} \in \mathbb{W}_{j,:}$}{
                \ccomment[blue]{Step 1.1: Separate inlier and outlier in each Macro-Block}\
                
                $\mathbb{W}^{in}, \mathbb{W}^{out}$ = \texttt{sep\_in\_out}($\mathbb{W}_{j, MaB})$
                
                \ccomment[blue]{Step 1.2: Quantize Inliers to lower precision}\
                
                $\mathbb{Q}^{in}, I_{sf}$ = \texttt{InlierQuantization}($\mathbb{W}^{in}$)
                
                \For{$\mathbb{W}_{j,\mu B} \in \mathbb{W}_{j,MaB}$}{
                    \ccomment[blue]{Step 2.0: Count Number of Outliers in a Micro-Block}\
                    
                    \textit{n} = \texttt{min}($B_{\mu}/2$, \texttt{NumOutliers}($\mathbb{W}^{out}_{\mu B}$))
                    
                    \ccomment[blue]{Step 2.1: Initialize Inlier Index List}\
                    
                    M  $\leftarrow$ \{\} \\
                    \ccomment[blue]{Step 2.2: Identify n least Important Inlier Position}\\                   
                    \For{n iterations}{
                    
                        $p = argmin_{p \in \mathbb{W}^{in}_{\mu B}} w_p^2 /[\mathbb{H}^{-1}]_{pp}$

                        \ccomment[blue]{Step 2.3: Prune least important Inliers}\
                        
                        $w_p \leftarrow 0$
                        
                        \ccomment[blue]{Step 2.4: Update M with the location of $w_p$}\
                        
                        $M \leftarrow M + \{p\}$                        
                    }
                    
                    \ccomment[blue]{Step 2.5: Quantize Outliers to  higher precision}\
                    
                    $\mathbb{Q}^{out}, O_{sf}$ = \texttt{OutlierQuant}($\mathbb{W}^{out}_{\mu B}, I_{sf}$)

                    \ccomment[blue]{Step 3.0: Distribute LSB Outlier Bits to Sparse Inlier Indices}\
                    
                    perm $\leftarrow$ \texttt{DistributeOutlierBits}($\mathbb{Q}^{out}$, $M$)
                }

            $\mathbb{Q}_{j,MaB} = \mathbb{Q}^{in} + \mathbb{Q}^{out}$
                
            }
            
            \ccomment[blue]{Step 3.1: Quantization Error}\
            
            $\mathbb{E}_{(j-i), : } = (\mathbb{W}_{j, : } - \mathbb{Q}_{j, :}) / [\mathbb{H}^{-1}]_{jj} $
            
            \ccomment[blue]{Step 3.2: Update weights in rB to compensate quantization error}\
            
            $\mathbb{W}_{j:(i+rB),:} = \mathbb{W}_{j:(i+rB),:} - \mathbb{E}_{(j-i), : }\cdot\mathbb{H}^{-1}_{j:(i+rB), j}$
        }
        \ccomment[blue]{Step 3.3: Update remaining weights after a row block is quantized}\
        
        $\mathbb{W}_{(i+rB):,:} = \mathbb{W}_{(i+rB):,:} - \mathbb{E}\cdot\mathbb{H}^{-1}_{(i+rB):, i:(i+rB)}$
    }
\end{algorithm}

%% file: sections/hw_methodology.tex
\section{MicroScopiQ Accelerator Architecture}
\label{sec:hardware}

\begin{figure}[t]
    \centering
    \includegraphics[width=\columnwidth, keepaspectratio]{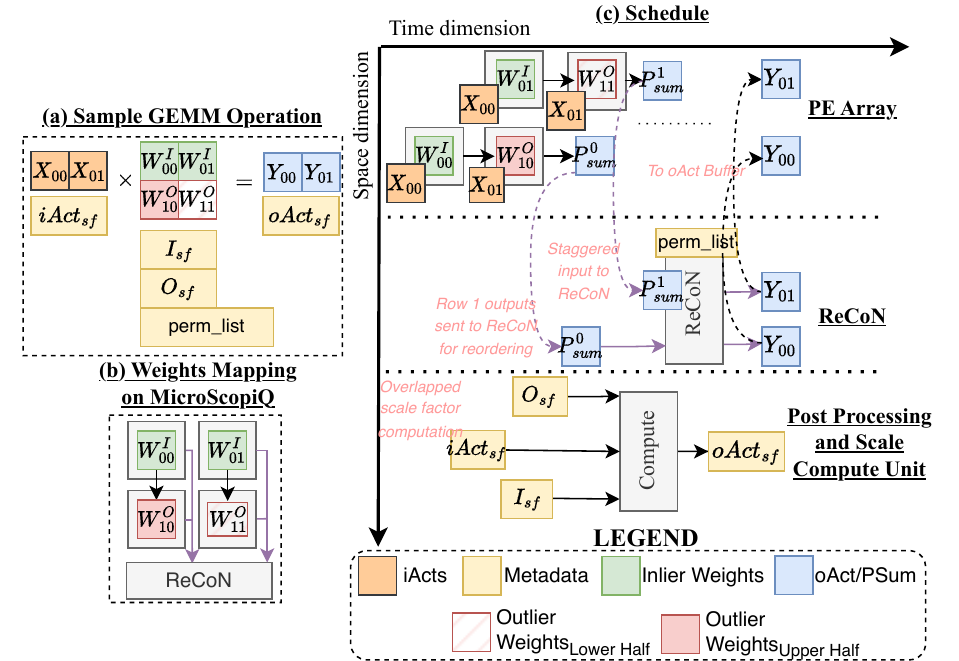}
        \vspace{-6mm}
        \caption{Example scheduling of a GEMM operation on a $2\times2$ MicroScopiQ accelerator.}
        \label{fig:memory_org_tiling}
\end{figure}

\subsection{Architecture Overview}
In \autoref{fig:hw_architecture}, we depict the MicroScopiQ accelerator architecture, integrated into a standard weight stationary systolic array. The accelerator is optimized for throughput and all compute units (PE array, ReCoN etc.) are internally \textbf{pipelined} with interleaved pipeline stages. \autoref{fig:memory_layout} illustrates MicroScopiQ's interaction with on-chip and off-chip memories. Following prior work \cite{keller202395, fang2024anda, dai2021vs}, off-chip memory is modeled as HBM2 with a 256 GB/s bandwidth. The off-chip memory layout \cite{guo2023olive, zadeh2020gobo} of a MicroScopiQ-quantized tensor, shown in \autoref{fig:memory_layout}, consists of two sections: quantized weights and metadata. Following prior work \cite{guo2023olive, guo2022ant, ramachandran2024algorithm} a two-level on-chip memory hierarchy is employed: 2MB L2 global SRAM loads data from DRAM and transfers it to MicroScopiQ buffers (\cref{sec:on-chip}) via an Open Compute Protocol OCP-SRAM interface \cite{sumbul2022system} with a bandwidth of 64 GB/s.

Through the numbered steps in \autoref{fig:hw_architecture}, we explain the computational flow of the MicroScopiQ accelerator. And in \autoref{fig:memory_org_tiling}(a-c), we demonstrate the mapping and schedule for a sample $2\times2$ GEMM operation with inlier and outlier weights, following the same computation flow.

Assume a quantized FM with outliers distributed within a $\mu$B and corresponding metadata calculated offline. 

\circled{1} One $\mu$B or multiple $\mu$Bs of weights are mapped to each PE row as typically $B_{\mu} \le$ $\#$ of columns in PE array \cite{jouppi2017datacenter}. Each PE in a row either receives a weight at high-precision (e.g. 4-bit) or multiple packed weights at low-precision (e.g. two 2-bits).  

\circled{2} PE row 0 receives quantized input activation (iAct) from left and input accumulation (iAcc) from top. The PEs in row 0 perform multiplication of the stationary weights with the iActs, regardless of outlier/inlier weights. During accumulation, assuming the $\mu$Bs mapped to PE row 0 do not contain outliers (see \autoref{fig:memory_org_tiling}(c)), the PEs accumulate the computed multiplication result with iAcc and direct the partial sums to PE row 1. 

\circled{3} PE row 1 receives partial sums from the top and similarly performs INT multiplication between weights and iAct. As shown in the scheduling figure, in parallel, the output scale factors can be calculated (\cref{sec:post}). 

\circled{4} The mapped $\mu$Bs to PE row 1 have outliers, the controller directs all PE outputs to ReCoN. PE row 1 offloads the accumulation to ReCoN (shown by dotted lines in \autoref{fig:memory_org_tiling}(c)). This is needed because the outlier will have its two halves distributed in different locations and mapped to different PEs. While the PEs in row 0 with inlier weights can accumulate the partial sum output, the PEs in row 1 with the \texttt{Upper} and \texttt{Lower} outlier halves cannot compute the outlier partial sum output. After receiving outputs from row 1, ReCoN reorders and calculates FP-outlier partial sum.  

\circled{5} The controller signals PE row 2 to expect input partial sums from ReCoN and not from the previous row. If PE row 1 is the last row as in \autoref{fig:memory_org_tiling}(c), the oActs are directed to the oAct buffer. 

\circled{6} oActs are post-processed (scaled and quantized). The post-processing of oActs does not require all the oActs to be computed. The post processing operations can be conducted as and when the oActs are generated and overlapped with the rest of the computation so as to hide computation and memory access latency overhead.

\vspace{-2mm}
\subsection{On-chip Storage and Control}
\label{sec:on-chip}
\noindent\textbf{Weight/Activation Organization. }The input activations (iActs) are stored in the iAct buffer as 8-bit INTs. Lower-precision iActs ($<$8-bit) are supported through sign-extension. The same process is followed for oActs (output activations). The weight buffer stores weights at 4-bit granularity. At lower precision (2-bits), each buffer location simultaneously stores two 2-bit weights. The \textbf{MODE} signal from the controller delineates the bit format of weights in the weight buffer i.e, one 4-bit weight or two 2-bit weights.

\noindent\textbf{Instruction Buffer (IB). }It stores all the metadata required for inference of MicroScopiQ quantized FMs. Particularly, the IB stores the outlier distribution permutation list i.e., configurations for ReCoN and scale factors. 

\noindent\textbf{MicroScopiQ Controller. }The controller generates appropriate control and signals to all units. It is functionally very similar to standard systolic array controllers \cite{sharma2018bit}, however, with added functionality to support specific features of ReCoN (arbitration of simultaneous access by multiple rows), multi-precision PEs, post-processing unit. Furthermore, unlike traditional controllers, the MicroScopiQ controller exerts fine-grained control using handshaking signals on the streaming of iActs, iAccs into the PE array to account for the increased pipeline depth through ReCoN for rows with outliers. The controller also generates OCP-SRAM interface control signals to manage data flow between the buffers and the L2 global SRAM.

\vspace{-2mm}
\subsection{Multi-Precision PE Array}
\label{sec:mixed_precision}
\begin{figure}[t]
\vspace{-0.5mm} 
    \centering
    \includegraphics[width=\columnwidth, keepaspectratio]{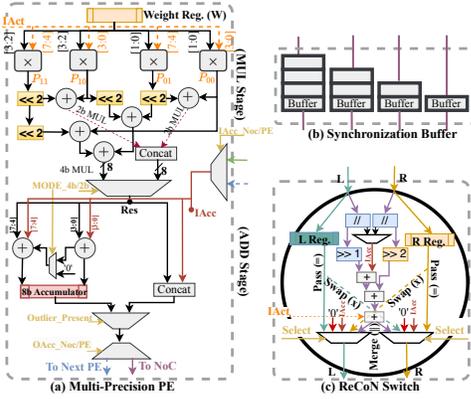}
        \vspace{-8mm}
        \caption{Microarchitecture of (a) Multi-precision PE, (b) Synchronization buffer and c) ReCoN switch.}
        \label{fig:hw_components}

\end{figure}

\textbf{Multi-Precision. }We depict the multi-precision MicroScopiQ PE in \autoref{fig:hw_components}(a). Existing multi-precision accelerators \cite{guo2023olive, sharma2018bit, guo2022ant} typically follow a bottom-up approach, employing low-precision PEs and grouping neighboring PEs to support higher precision. This sacrifices throughput for multi-precision support, as multiple PE columns are required to perform a single MAC operation. This reduces parallelism and increases latency. We adopt a different strategy via the \textbf{MODE} signal for multi-precision support by mapping multiple weights that share the same rightbound iAct to the same PE at lower precision (2-bits) or a single weight at higher precision (4-bits). The two 2-bit weights mapped to the same PE in MicroScopiQ are the weights that would have been mapped to different columns within the same row in existing multi-precision accelerators. At high-precision operation we utilize available parallelism from the PE array while at lower-precision we increase throughput by the parallel evaluation of multiple partial sums.

\noindent
\textbf{Multiplication Stage. }Inspired by \cite{liu2023high, umuroglu2018bismo} we present a multiplier-tree architecture for multi-precision INT multiplication. The weights and iAct are partitioned across the four $4$-bit $\times$ $2$-bit multipliers to calculate partial sums ($P_{00}, P_{01}, P_{10}, P_{11}$) as described in \autoref{fig:hw_components}(a). Based on the \textbf{MODE} signal the partial sums above are combined using a combination of adders and shifters to provide the result (\texttt{Res}) of different bit-precision weights with iAct as follows,
\vspace{-2mm}
\begin{equation}
\small
    \texttt{Res} = 
    \begin{cases}
    \{(P_{11} << 2 + P_{10}), (P_{01}<<2 + P_{00}) \}, \textbf{MODE}_{2b}\\
     (P_{11} << 4 + P_{00}) + (P_{01} << 2 + P_{10} << 2),  \textbf{MODE}_{4b}
    \end{cases}
\end{equation}
\vspace{-3mm}

\noindent
\textbf{Accumulation Stage. }It receives the MUL-stage output (\texttt{Res}) and also iAcc from either ReCoN or the previous PE row. The controller signals the ADD stage via \textbf{Outlier\_Present} to modify accumulation behavior based on whether inlier/outlier weights are present in the PE. If the \texttt{Res} corresponds to a multiplication of inlier weight with iAct, then the corresponding \texttt{Res} is added with iAcc using two adders with a multiplexer that enables multi-precision by propagating the carry from one adder to other. In low precision \textbf{MODE}, the two adders work independently to calculate partial sums in parallel. For high precision \textbf{MODE}, both adders work together to produce a single accumulation result with carry propagation through the multiplexer. On the other hand, if \texttt{Res} corresponds to multiplication of either of the outlier halves with iAct, the actual outlier accumulation is offloaded to ReCoN by concatenating \texttt{Res} and iAcc\footnote{the concatenated output is notated as $O_{\texttt{Upper},\texttt{Res}}$ or $O_{\texttt{Lower},\texttt{Res}}$.}. This is done to prevent incorrect outlier partial sum calculation. The controller directs partial sum outputs to ReCoN or next PE row based on presence of outliers, via the \textbf{OAcc\_NoC/PE} signal.

\vspace{-2mm}
\subsection{Redistribution and Coordination NoC}
\label{sec:recon}
The Redistribution and Coordination NoC (ReCoN) is a multistage butterfly NoC in the MicroScopiQ accelerator \textbf{time-multiplexed} and shared across PE rows (see \cref{sec:recon_analysis}). This is a more \textbf{cost effective} way of handling outliers compared to OliVe and GOBO \cite{guo2023olive, zadeh2020gobo} that handle outliers within the PE. Since \textbf{PEs are large in number} OliVe, GOBO will incur larger costs compared to MicroScopiQ.

\noindent
\textbf{ReCoN topology. }ReCoN is composed of $n(\log_2(n) + 1)$ \{2-input, 2-output\} ReCoN switches, in a multistage butterfly NoC topology \cite{krishna2014smart}. The input and output stages of ReCoN also employ \{2-input, 2-output\} switches, with a dedicated switch for each column receiving (transmitting) partial sums into (from) one of the two input (output) ports. The other port of the input and output stages of each switch is tied to $0$. Each ReCoN stage also receives the same iActs used by the PEs to compute the partials sums. As we shall show later, this is done to facilitate FP-outlier's hidden-bit processing.

\begin{figure}[t]
    \centering
    \includegraphics[width=\columnwidth, keepaspectratio]{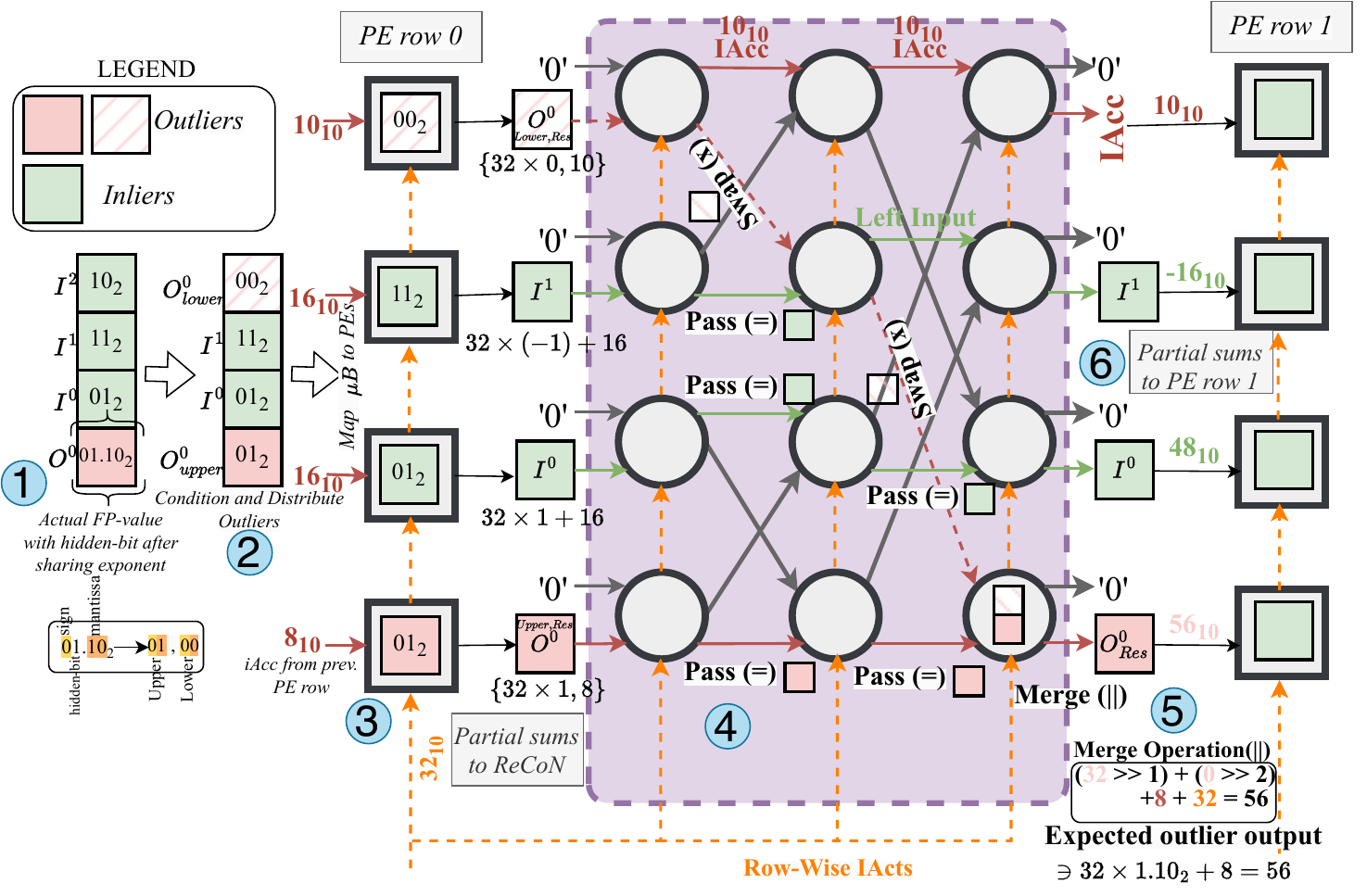}
        \vspace{-8mm}
        \caption{End-to-end example illustrating the working of MicroScopiQ with a $4\times4$ PE array (only two rows shown).}
        \label{fig:noc_example}
        \vspace{1pt}
\end{figure}

\noindent
\textbf{ReCoN Input Interface. }ReCoN is shared and time-multiplexed across all PE rows, and at a particular instant many rows might require access to it. At the inputs, the column-wise arbiters guided by control signals from the controller, coordinate and resolve contention between different PE rows and ensure a fair scheduling (for simplicity in \autoref{fig:hw_architecture}, we conceptually depict this operation as an N-input multiplexer). Due to the skewed data flow of systolic array computations \cite{jouppi2017datacenter, ramachandran2024algorithm}, partial sums from different columns of the same row do not arrive at the ReCoN input at the same cycle. A synchronization buffer (\autoref{fig:hw_components}(c)) is positioned between ReCoN and column-wise multiplexers which propagates the partial sums through sets of buffers of different lengths (column producing the partial sum fastest has the most buffers), synchronizing the arrival of inputs from different columns. The synchronization buffer sends an ACK signal to the PE row whose values have been accepted so that it can progress in its computation and the other simultaneously contesting rows will be on hold and acknowledged in the next cycle. When there are N simultaneously contesting rows, the row that is acknowledged the last adds an N-1 cycle latency to its processing. As we show in \autoref{fig:outlier_group}(b), \autoref{fig:recon}(a) the num. of conflicts in accessing ReCoN are minimal and if latency is a critical factor multiple ReCoN units can be used to minimize conflicts.

\noindent
\textbf{ReCoN switch. }It is a \{2-input, 2-output\} switch (\autoref{fig:hw_components}(c)) that performs three major operations based on a 3-bit configuration:

\begin{itemize}[align=right,itemindent=2em,labelsep=2pt,labelwidth=1em,leftmargin=0pt,nosep]
  \item \texttt{Pass} ($=$): Passes the input from left (right) port to left (right) output port. 
  
  \item \texttt{Swap} ($\times$): Directs the input from the left (right) input to the right (left) output, with the opposite output port receiving the right (left) port input, 0 or iAcc (see\cref{sec:recon_action}).
  
  \item \texttt{Merge} ($||$): This function is triggered when a switch receives $O_{\texttt{Upper},\texttt{Res}}$ and $O_{\texttt{Lower},\texttt{Res}}$ at its left and right ports, respectively. The switch separates the \texttt{Res} from iAcc (denoted as // in \autoref{fig:hw_architecture}) from both inputs and selects the \texttt{Upper} result's iAcc, since it is the correct iAcc for accumulation ($O_{\texttt{Lower},\texttt{Res}}$ iAcc is selected during \texttt{Swap}). The \texttt{Upper} and \texttt{Lower} halves of outlier weight's magnitude bits are actually mantissa bits ($<1$) of MX-FP which the PE treat as INTs. Therefore the \texttt{Res} of $O_{\texttt{Upper},\texttt{Res}}$ and $O_{\texttt{Lower},\texttt{Res}}$ are shifted $1\times$ and $2\times$ respectively (not shown in figure, but internally handles multi-precision \textbf{MODE}) to account for the decimal position of mantissa in the MX-FP format. The shifted \texttt{Res} are then accumulated with the corresponding iAcc. FP-formats also have a hidden-bit ($1.0$) \cite{wang2018training, ramachandran2024algorithm}; to account for the contribution of the hidden bit to the outlier partial sum we also add iAct to the accumulation above.
  
\end{itemize}

  
\noindent
\textbf{ReCoN Output Interface. }The reordered partial sums are then routed to the subsequent row in the PE array or if the input is from the last PE row, to the oAct buffer. Since ReCoN is pipelined internally and with the rest of the PE array, reordered and processed partial sums are produced every cycle once the pipeline depth (=number of ReCoN stages) is filled.

\vspace{-5mm}
\subsection{Post Processing and Scale Compute Unit}
\label{sec:post}
The shared output scale factor for $B_M$=128 is calculated as $ {oAct_{sf}} = {O_{sf} + iAct_{sf}}$ (See expansion of $O_{sf}$ in \cref{sec:inlier_outlier}.). The calculation of $oAct_{sf}$ (with simple adders and subtractors) is independent of the processing done in the PE array. Therefore, we overlap the computation of $oAct_{sf}$ with the processing of oActs to efficiently hide the computation latency similar to prior works \cite{dai2021vs, ramachandran2024algorithm}. Since we maintain different scales for inliers and outliers (see \cref{sec:inlier_outlier}), and the oAct scale factor is calculated based on the outlier scale factor ($O_{sf}$) (we found this to result in least quantization error of oActs), the oActs which are generated through computation with only inlier weights (identified through the in/out control signal) are shifted by the shift value depicted in the ``Post Processing and Scale Compute Unit'' in \autoref{fig:hw_architecture} to ensure conformity with the final scale factor. Finally, the oActs are scaled with the computed output scale factor through a simple right shift (since it is a power-of-two scale factor, division can be implemented through right shift operation) and quantized to MX-INT-(4/8)$_{128}$ before being sent to external memory or routed back to the iAct buffer for computation with the next layer's weights. Similar to prior works \cite{ramachandran2024algorithm}, \cite{keller202395} the post processing unit is also responsible for handling all non-linear operations.
\begin{figure}[t]
    \centering
    \includegraphics[width=\columnwidth, keepaspectratio]{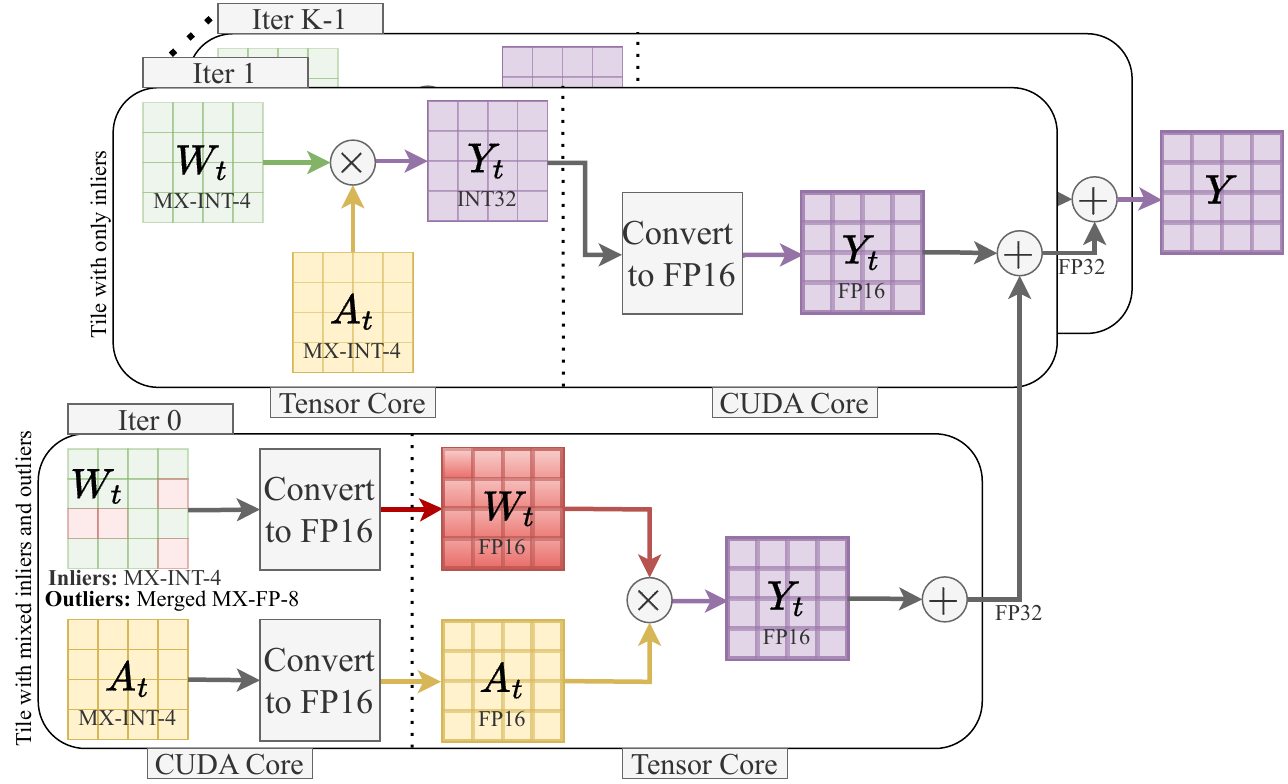}
        \vspace{-4mm}
        \caption{Illustration of W4A4 MicroScopiQ GEMM on a GPU.}
        \label{fig:gpu_cuda}
\end{figure}

\input{tables/llm_results}

\subsection{MicroScopiQ Walkthrough Example}
\label{sec:recon_action}
In this section, we discuss an end-to-end example that concretely showcases the working of MicroScopiQ. In the example shown in \autoref{fig:noc_example}, we employ a $4\times4$ systolic array (2 rows shown) with PE row 0 having one outlier and three inliers in the $\mu$B and PE row 1 having all inlier weights. Note: Not all rows require access to ReCoN, only rows that have one or more outlier weights (row 0) in the PE require access. We assume $\mu$B=4, inlier and outlier quantization formats of MX-INT-2$_{128}$ and MX-FP-4$_{4,4}$, respectively. 

\circled{1} To demonstrate that outlier partial sums processed through ReCoN results in accurate FP-outlier partial sum, we first show a $\mu B$ with one outlier and its actual FP-value after sharing of $\mu X$, and having a value of $1.5_{10}$. If there were no outlier distribution, the actual partial sum from an ``outlier-specific PE'' for the shown iAct=$32_{10}$ and iAcc=$8_{10}$ will be $56_{10}$. However as described in \cref{sec:outlier_pruning}, we distribute outliers within a $\mu$B into \texttt{Upper} and \texttt{Lower} halves. Note: The hidden-bit is implicit and is only shown for demonstration. 

\circled{2}The $\mu$B after distributing the outlier is mapped to PE row 0. 

\circled{3} PE row 0 computes partial sum outputs for inliers in columns 1 and 2, while it outputs $O_{\texttt{Upper},\texttt{Res}}$ and $O_{\texttt{Lower},\texttt{Res}}$ for the other two columns that have outliers. The output of PE row 0 is routed through ReCoN. In ReCoN, the inliers are \textbf{passed} down at all stages, since they are already present in their respective columns.

\circled{4}The level 1 switch at column 3, executes a \textbf{swap} operation and redirects the lower half outlier output towards the upper half while simultaneously sending iAcc through the other port. This iAcc corresponds to the iAcc input to column 3 PE in row 0. This is functionally correct because, the lower half of outlier is distributed in this column by zeroing out/pruning the inlier at that location, hence the partial sum output of that PE to the corresponding PE in the next row should only be iAcc since the original weight at that location is zero. Similarly in level 2 of ReCoN, the switch that receives the $O_{\texttt{Lower},\texttt{Res}}$ from level 1 again executes a \textbf{swap} operation. Finally in level 3 the outlier output halves are merged together to realize $O^0_{Res}$, while all other switches execute pass. This merge operation results in a $O^0_{Res}$ that matches the expected output. 

\circled{5}, \circled{6} The partial sums from ReCoN are then sent to PE row 1 which then perform MAC operations and sends it to the subsequent row since there are no outlier weights. Similarly all rows execute operations until the final partial sum is calculated.

\section{MicroScopiQ Integration in GPUs}
\label{sec:gpu}
In this section, we demonstrate the integration of MicroScopiQ into GPUs via SW (\cref{sec:kernel}) and HW (\cref{sec:tensor}) modifications. For a matrix multiplication problem of size $M \times N \times K$, each GPU thread block \cite{lin2024qserve, zhao2024atom} is responsible for computing a $T_m \times T_n$ output tile. This computation is performed iteratively along the $K$ dimension (\autoref{fig:gpu_cuda}), which is referred to as the main loop. MicroScopiQ quantized models face acceleration challenges on GPUs due to: (a) high-cost pointer arithmetic from distributed outliers and (b) tensor cores' inability to co-issue INT and FP operations \cite{nvidia2017volta}, preventing direct acceleration of tiles. Our SW, HW optimizations address these issues to maximize GPU performance.

\vspace{-1mm}
\subsection{SW: Kernel-Level Optimizations}
\label{sec:kernel}
\textbf{Register Caching. }We implement a virtual caching layer \cite{ben2016fast} to efficiently distribute outliers within a warp. Instead of each thread block loading and merging outliers from shared memory, we use \texttt{shfl\_sync(m, r, t)} \cite{ben2016fast} for intra-warp communication. This primitive allows a thread to share its register \texttt{r} while reading the value from thread \texttt{t} within the same warp, using \texttt{m} as the thread selector mask. With $B_{\mu}=8$, each warp (32 threads) has 4 $\mu B$s, allowing efficient intra-warp outlier merging based on the permutation list.

\noindent
\textbf{GEMM on Tensor Cores. }Due to MicroScopiQ’s MX-FP format for outliers and MX-INT for inliers, mixed tiles after outlier merging (iteration 0 in \autoref{fig:gpu_cuda}) must first be dequantized to FP16 before GEMM execution. In contrast, inlier-only tiles (iteration 1) can leverage 4-bit INT tensor cores for efficient acceleration but still require INT32-to-FP16 dequantization for accumulation along the K-dimension (\mbox{\autoref{fig:gpu_cuda}}) with other tiles. We implement this block-level dynamic decision to deliver maximum performance. However, the need for repeated dequantization and lack of multi-precision support is a significant bottleneck on GPUs.       

\subsection{HW: Potential Tensor Core Modification}
\label{sec:tensor}
Following prior modeling \cite{guo2023olive, raihan2019modeling}, each tensor core \cite{nvidia2017volta, abdelkhalik2022demystifying} performs 16-bit FEDPs (four-element dot products). Therefore, each tensor core at 4-bits can conduct 16EDPs. Efficiently accelerating MicroScopiQ requires simultaneous INT+FP support within the tensor core. Following the functionality of ReCoN in \cref{sec:recon}, each 16EDP operation requires a variable right shifter (Inliers: $>>0$, Outlier Upper Half: $>>1$, Outlier Lower Half: $>>2$) to account for the FP mantissa for outlier products. With typical GPU die sizes (RTX 2080 Ti: 754 $mm^2$), adding a shifter has negligible overhead ($\sim 0.1\%$).

%% file: tables/llm_results.tex
\begin{table*}[t]
    \centering
    \caption{Quantization Results for LLMs. We report WikiText2 perplexity numbers (lower the better).}
    \vspace{-5pt}
    \resizebox{0.95\linewidth}{!}{%
    \begin{tabular}{lc|cc|ccc|cc|c|cc}
    \Xhline{2\arrayrulewidth}
    \rowcolor[HTML]{E0E0E0}
    &   & \multicolumn{2}{c|}{\textbf{OPT}\cite{zhang2022opt}} & \multicolumn{3}{c|}{\textbf{LLaMA-2}\cite{touvron2023llama}} &\multicolumn{2}{c|}{\textbf{LLaMA-3}\cite{meta2024introducing}} & \textbf{Mixtral} \cite{jiang2024mixtral} & \multicolumn{2}{c}{\textbf{Phi-3}\cite{abdin2024phi}} \\
    \cline{3-12}
    \rowcolor[HTML]{E0E0E0}
    \textbf{Method} & \textbf{W/A}  & \textbf{6.7B} & \textbf{175B} & \textbf{7B} & \textbf{13B} & \textbf{70B} & \textbf{8B} & \textbf{70B} & \textbf{8x7B} & \textbf{3.8B} & \textbf{14B}   \\
    \Xhline{2\arrayrulewidth}
    \rowcolor[HTML]{D9EAF5}
    Baseline & 16/16 & 10.86 & 8.34 & 5.47 & 4.83 & 3.31 & 6.13 & 2.85 & 3.84 & 6.33 & 4.31   \\
    \Xhline{1\arrayrulewidth}
    OliVe \cite{guo2022ant} & 4/16 & 12.20 & 9.09 & 11.52 & 9.34 & 7.23 & 10.29 & 5.65 & 6.19 & 8.57 & 7.81  \\
    GOBO \cite{zadeh2020gobo} & 4/16 & 10.97 & 8.71 & 5.79 & 5.03 & 3.45 & 7.11 & 3.53 & 4.22 & 6.64 & 4.78  \\
    GPTQ \cite{frantar2022gptq} & 4/16 & 11.12 & 9.09 & 6.23 & 5.58 & 4.28 & 8.12 & 3.75 & 4.68 & 7.17 & 5.13  \\
    AWQ \cite{lin2024awq} & 4/16 & 10.97 & 8.74 & 5.82 & 5.19 & 4.08 & 7.96 & 3.58 & 4.36 & 6.72 & 4.99   \\
    OmniQuant \cite{shao2023omniquant} & 4/16 & 10.96 & 8.72 & 5.74 & \textbf{5.02} & 3.47 &7.09 & 3.46 & 4.19 & 6.67 & 4.82   \\
    \rowcolor[HTML]{D5E8D4}
    \textbf{MicroScopiQ (Ours)} & \textbf{4/16} & \textbf{10.91} & \textbf{8.62} & \textbf{5.65} & \textbf{5.02} & \textbf{3.42} & \textbf{6.89} & \textbf{3.25} & \textbf{4.07} & \textbf{6.61} & \textbf{4.70}  \\
    \Xhline{1\arrayrulewidth}
    OliVe \cite{guo2023olive} & 4/4 & 55.44 & 14.17 & 19.28  & 14.96 & 13.59 & 27.65 & 9.34 & 23.53 & 17.63 & 15.29\\
    OmniQuant \cite{shao2023omniquant} & 4/4 & 11.61 & 9.88 & 11.47 & 8.32 & 5.41 & 10.21 & 5.30 & 5.98 & 8.21 & 6.40   \\
    SmoothQuant \cite{xiao2023smoothquant} & 4/4 & 19.54 & 17.62 & 20.47 & 15.63 & 17.62 & 29.54 & 19.32 & 37.54 & 18.11 & 15.39   \\
    Atom \cite{zhao2024atom} & 4/4 & 11.15 & 9.02 & 6.16 & 6.12 & 5.20 & \textbf{8.12} & 4.69 & 5.35 & 7.59 & 5.95   \\
    \rowcolor[HTML]{D5E8D4}
    \textbf{MicroScopiQ (Ours)} & \textbf{4/4} & \textbf{10.97} & \textbf{8.95} & \textbf{6.11} & \textbf{5.57} & \textbf{4.48} & \textbf{8.12} & \textbf{4.65} & \textbf{5.03} & \textbf{6.95} & \textbf{5.41}   \\
    \Xhline{1\arrayrulewidth}
    OmniQuant \cite{shao2023omniquant} & 2/16 & 11.61 & 9.66 & 9.62 & 7.56 & 6.11 & 9.13 & 6.17 & 6.02 & 7.09 & 6.28   \\
    SDQ \cite{jeong2024sdq} & 2/16 & 12.09 & 10.04 & 10.47 & 8.09 & 6.98 & 10.54 & 6.93 & 7.62 & 7.39 & 6.92   \\
    \rowcolor[HTML]{D5E8D4}
    \textbf{MicroScopiQ (Ours)} & \textbf{2/16} & \textbf{11.51} & \textbf{9.42} & \textbf{8.43} & \textbf{7.06} & \textbf{6.01} & \textbf{8.97} & \textbf{5.91} & \textbf{6.02} & \textbf{7.16} & \textbf{6.03}   \\
    \Xhline{1\arrayrulewidth}
    OmniQuant \cite{shao2023omniquant} & 2/8 & 11.99 & 10.23 & 9.62 & 8.92 & 6.83 & 9.39 & 6.59 & 6.29 & 7.95 & 7.37   \\
    Atom \cite{zhao2024atom} & 2/8 & 11.95 & 10.13 & 9.23 & 8.54 & \textbf{6.33} & 9.13 & 6.35 & \textbf{6.14} & 7.46 & 7.29   \\
    \rowcolor[HTML]{D5E8D4} 
    \textbf{MicroScopiQ (Ours)} & \textbf{2/8} &\textbf{11.77} & \textbf{9.98} & \textbf{9.06} & \textbf{8.06} & \textbf{6.33} & \textbf{9.08} & \textbf{6.02} & 6.17 & \textbf{7.38} & \textbf{6.82}   \\
    \Xhline{2\arrayrulewidth}
    \end{tabular}}
    \label{tab:llm_results}
    \vspace{-3mm}
\end{table*}

%% file: sections/results.tex
\section{Experimental Evaluations}
\label{sec:results}
\subsection[Experimental Setup]{Experimental Setup\footnote{We denote each quantization configuration by its $bb$ to enhance readability and facilitate clear comparisons.}}
\label{sec:results_intro}
\noindent
\textbf{Models and Datasets. }We evaluate on OPT \cite{zhang2022opt}, LLaMA2 \cite{touvron2023llama} LLaMA3 \cite{meta2024introducing}, Phi-3 (SLM) \cite{abdin2024phi} and Mixtral (MoE) \cite{jiang2024mixtral} LLM model families. The VLMs evaluated are OpenFlamingo \cite{awadalla2023openflamingo}, VILA \cite{lin2024vila} and LlaVa \cite{liu2024visual}. We use a calibration dataset consisting of 256 random samples from the PILE dataset \cite{gao2020pile}, so as to not overfit to any particular downstream domain \cite{lin2024awq}. We compare different LLMs based on perplexity (PPL $\downarrow$) on the WikiText2 \cite{merity2016pointer} dataset and benchmark accuracy on 6 different tasks BoolQ \cite{clark2019boolq}, HellaSwag \cite{zellers2019hellaswag}, PIQA \cite{bisk2020piqa}, ARC-c \cite{clark2018think}, MMLU \cite{hendrycks2020measuring} and Winogrande \cite{sakaguchi2021winogrande}. 

Similarly, we evaluate VLMs on 5 vision-language benchmarks: COCO captioning \cite{chen2015microsoft}, VQAv2 \cite{goyal2017making}, VizWiz \cite{gurari2018vizwiz}, TextVQA \cite{singh2019towards}, and GQA \cite{hudson2019gqa}. Additionally, to demonstrate generalizability we also evaluate Convolutional Neural Network (CNN) baselines-ResNet50, VGG16 and State Space Model (SSM) baselines-VMamba \cite{liu2024vmamba}, Vim \cite{zhu2024vision}. For these models we employ a calibration dataset of 64 samples from the Imagenet training set \cite{ramachandran2025ouromamba}.

\noindent
\textbf{Algorithm Implementation. }We implement the MicroScopiQ Quantization Framework in PyTorch \cite{paszke2019pytorch}. All FMs are quantized using a single NVIDIA H100 GPU. The complete quantization process' runtime ranges between 30 mins.--9 hours depending on model size (3.8B to 175B), which is at par with recent SoTA techniques \cite{frantar2022gptq, zhao2024atom}.

\noindent
\textbf{Algorithm Baselines. }We compare the performance of our MicroScopiQ quantization framework against existing SoTA quantization algorithms: OmniQuant \cite{shao2023omniquant}, AWQ \cite{lin2024awq}, GPTQ \cite{frantar2022gptq}, Atom \cite{zhao2024atom}, SDQ \cite{jeong2024sdq} and co-design techniques: OliVe \cite{guo2023olive}, GOBO \cite{zadeh2020gobo}. 

\noindent
\textbf{Accelerator Implementation. }The MicroScopiQ accelerator is implemented in Verilog RTL. We perform synthesis and PnR using Synopsys Design Compiler and Innovus with a TSMC 7nm technology library (See floorplan in \autoref{fig:die_recon}(a)) for area, power and latency calculations. All MicroscopiQ design variants attains a peak clock frequency of \textbf{1 GHz}. We use CACTI \cite{muralimanohar2009cacti} to estimate the area and power of on-chip memories. For end-to-end performance evaluation, we design a cycle accurate simulator based on DnnWeaver \cite{sharma2016high} and BitFusion \cite{sharma2018bit}, following previous works \cite{guo2023olive, ramachandran2024algorithm}.  

\noindent
\textbf{Accelerator Baselines. }We compare the MicroScopiQ accelerator against OliVe \cite{guo2023olive}, ANT \cite{guo2022ant}, GOBO \cite{zadeh2020gobo}, OLAccel \cite{park2018energy} and AdaptivFloat \cite{tambe2020algorithm} across area, power and performance metrics. To demonstrate the minimal overhead of ReCoN on NoC-based \textbf{real-world} industrial and academic accelerators, we also implement and compare against baseline MTIA \cite{firoozshahian2023mtia} and Eyeriss-v2 \cite{chen2019eyeriss}. For a fair comparison, we ensure all designs attain a clock frequency of 1GHz and have same memory hierarchy and bandwidth. DeepScale \cite{sarangi2021deepscaletool} is employed to scale all designs to the 7nm process.

\noindent
\textbf{GPU Implementation. }The optimized MicroScopiQ kernel is implemented in CUDA \mbox{\cite{zhao2024atom}} with a PyTorch frontend. We extend GPGPU-Sim \cite{bakhoda2009analyzing} and AccelSim \cite{khairy2018exploring} for Ampere GPUs to integrate and evaluate the modified tensor core. Energy estimation is performed using AccelWattch \cite{khairy2020accel}, GPUWattch \cite{leng2013gpuwattch}. Real GPU inference throughput follows the setup in \cite{zhao2024atom}, while simulated GPU results follow \cite{guo2023olive}.

\noindent
\textbf{GPU Baselines. }We compared system-level performance with FP16 TensorRT-LLM \cite{tensorrt} baseline and W4A4 Atom kernel \cite{zhao2024atom}.

\input{tables/additional_llm_results}
\subsection{LLM Quantization Results} 
\noindent \textbf{Weight-Only Quantization. }In \autoref{tab:llm_results}, we compare the WikiText2 PPL of different LLMs for W4A16 and W2A16 settings. MicroScopiQ consistently outperforms all the baselines across different models and quantization settings. At W4A16, \textbf{MicroScopiQ achieves near-lossless quantization performance}. Particularly at W2A16, the benefits of the MicroScopiQ method is evident, achieving \textbf{up to a 2.04 decrease in PPL score} compared to the baselines. This is due to MicroScopiQ's ability to quantize outliers at higher-precision and in FP-format to reduce quantization error. At W2A16, techniques like AWQ and OliVe have unacceptable accuracy degradations ($\geq1e4$ PPL, not depicted in table). MicroScopiQ outperforms the contemporary SDQ~\cite{jeong2023vegeta} at W2A16 with up to a \textbf{2.06} lower PPL score, owing to its flexible pruning+quantization strategy. In contrast, SDQ relies on a rigid N:M pattern, limiting its adaptability across model families. MicroScopiQ also outperforms the SoTA quantization framework OmniQuant across all models which requires $5-6\times$ \textbf{higher runtime than MicroScopiQ}. 


\begin{figure}[t]
    \centering\includegraphics[width=\columnwidth, keepaspectratio]{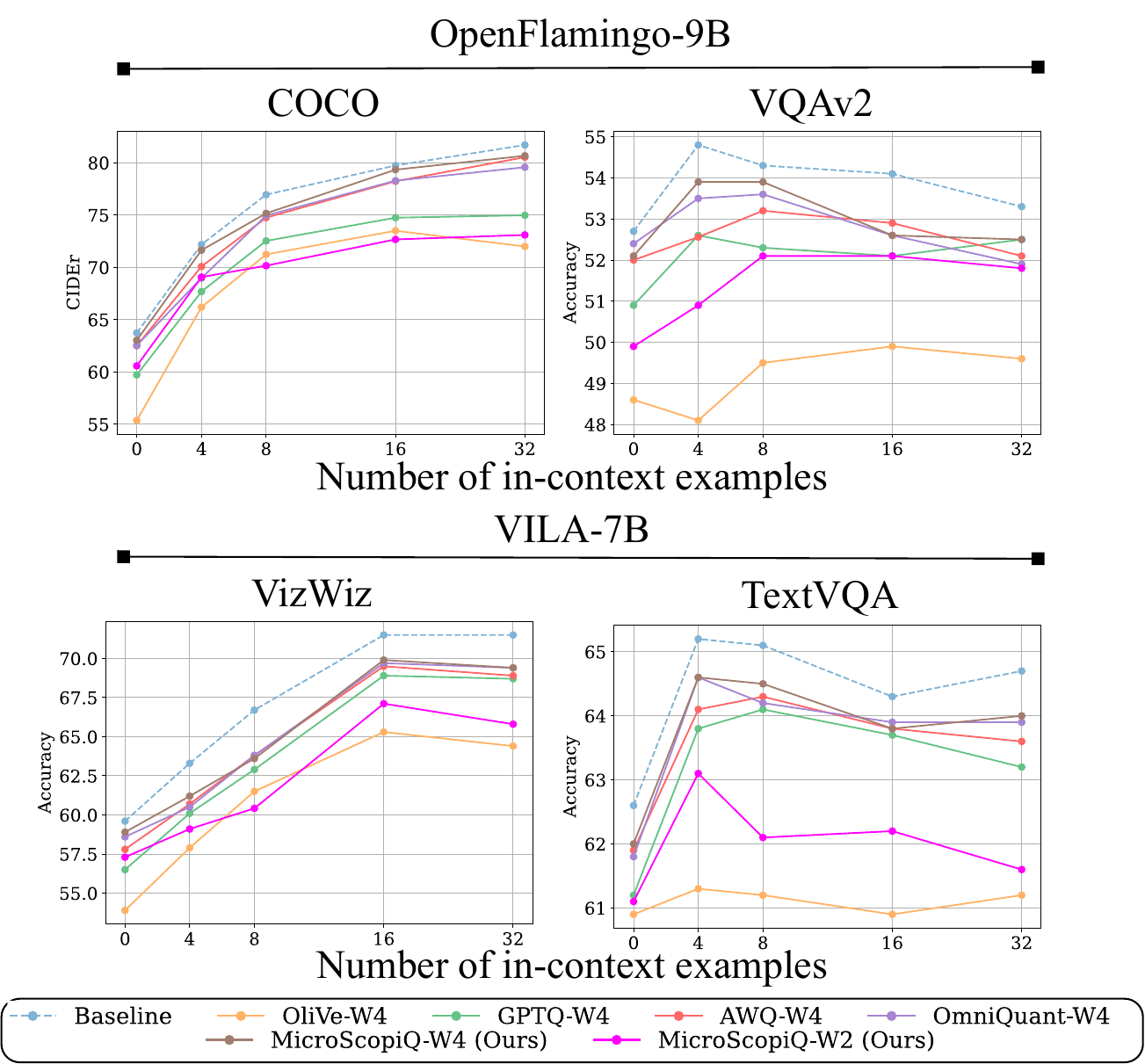}
        \vspace{-7mm}
        \caption{Weight-only quantization for VLMs, OpenFlamingo-9B \cite{awadalla2023openflamingo} on COCO, VQAv2 and VILA-7B \cite{lin2024vila} on VizWiz and TextVQA.}
        \label{fig:vlm_results}
\end{figure}

\begin{figure}[t]
    \centering\includegraphics[width=\columnwidth, keepaspectratio]{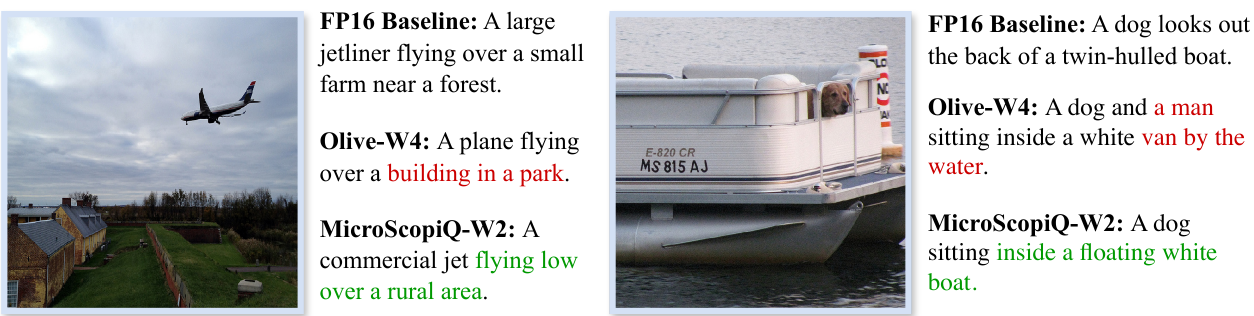}
        \vspace{-18pt}
        \caption{Qualitative results of OpenFlamingo-9B \cite{awadalla2023openflamingo} on 8-shot COCO captioning with weight-only quantization using OliVe (W4) and MicroScopiQ (W2). Text in \textcolor[HTML]{CC0000}{red} and \textcolor[HTML]{009900}{green} indicate incorrect and accurate captioning respectively.}

        \label{fig:vlm_prediction}
\end{figure}

\noindent \textbf{Weight-Activation Quantization. }Similarly, we compare the quantization performance of MicroScopiQ for W4A4 and W2A8 configurations in \autoref{tab:llm_results}. Due to the dynamic nature of activations, very rarely do techniques quantize outliers in activations directly \cite{zhao2024atom}. Instead, most techniques \cite{xiao2023smoothquant, shao2023omniquant} migrate the activation outliers to weights to enable simpler activation quantization. We borrow the migration strength ($\alpha$) hyper-parameter introduced in SmoothQuant \cite{xiao2023smoothquant} to migrate the difficulty of quantizing outliers in activations to weights. We find that prior techniques \cite{xiao2023smoothquant, shao2023omniquant} can only migrate up to 50\% of activation quantization difficulty to weights, i.e., $\alpha = 0.5$ before beginning to introduce higher weight quantization error. MicroScopiQ's robustness to higher presence of outliers in weights by effectively quantizing outliers at higher precision and identifying least important weights to prune, allows $\alpha$ as high as 0.7 i.e., migrating most of the activation outliers to weights. This enables simpler activation quantization with MX-INT-(4/8)$_{128}$ and absolving the need to handle outliers in activations. With $\alpha = 0.7$ for MicroScopiQ, $\alpha = 0.5$ for SmoothQuant and learned migration strength for OmniQuant in \autoref{tab:llm_results}, MicroScopiQ \textbf{outperforms all baselines with up to 7.4$\times$ lower PPL score} across quantization settings. Furthermore, MicroScopiQ outperforms a recent work Atom \cite{zhao2024atom} that performs activation quantization with up to a 0.33 lower PPL score.         

\noindent \textbf{Benchmark Accuracy. }In \autoref{fig:outlier_motivation}(b), we compare the zero-shot task accuracy of W2A16 MicroScopiQ against W416 OliVe for three different LLM benchmarks. MicroScopiQ at lower-precision than OliVe achieves $\geq$$\textbf{8\%}$ \textbf{higher accuracy} on the benchmarks compared to OliVe. OliVe shows poor accuracy due to its assumption on outlier locality, resulting in unintended outlier pruning. In \autoref{tab:llm_benchmarks} we also compare MicroScopiQ against OliVe and OmniQuant at W2A16 setting on 4 other LLM benchmarks. MicroScopiQ consistently outperforms all baselines by up to 9\% across all benchmarks.

\noindent
\textbf{Effective Bit Width. }For $bb=2, 4$, MicroScopiQ has an EBW of $2.36$ and $4.15$ bits respectively. Techniques like GOBO, have a very high EBW of $15.6$, $18.17$ bits. All other baselines employ software-managed metadata and have an EBW=$bb$.

\input{tables/non_transformer}
\begin{figure}[t]
    \centering\includegraphics[width=\columnwidth, keepaspectratio]{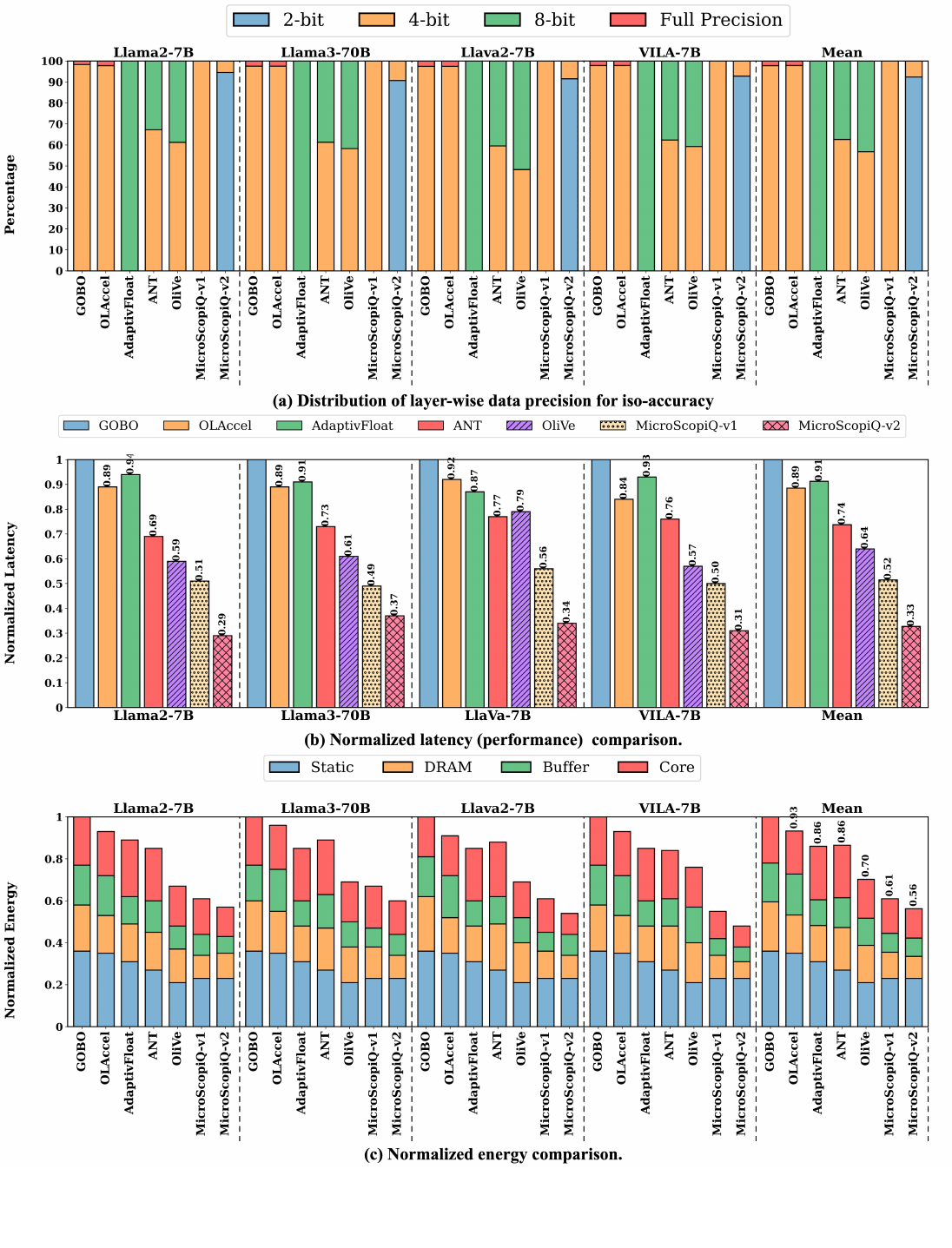}
        \vspace{-15mm}
        \caption{Iso-accuracy comparison of different accelerators. MicroScopiQ-v1 and -v2 are executed on the same accelerator and correspond to two different data-precision distributions.} 

        \label{fig:energy_perf}

\end{figure}
\subsection{VLM Quantization Results}
In \autoref{fig:vlm_results}, we compare 0-shot and multi-shot (4,8, 16, 32) weight-only quantization performance of OpenFlamingo-9B on the COCO captioning task and VILA-7B on VizWiZ, TextVQA benchmarks. At W4A16 quantization, MicroScopiQ consistently outperforms all baselines and achieves on average \textbf {less than $\textbf{1\%}$ accuracy drop} compared to the full-precision baseline, demonstrating the flexibility and widespread applicability of our method. Furthermore, at W2A16 quantization, MicroScopiQ achieves high accuracy ($<4\%$ accuracy drop), outperforming several W4A16 baselines.

We also quantitatively evaluate the performance of MicroScopiQ W2A16 quantized OpenFlamingo-9B VLM on the quality of generated captions for images drawn from the COCO captioning task in \autoref{fig:vlm_prediction}. The MicroScopiQ quantized model generates accurate captions and preserves overall word semantics compared to OliVe. For instance, in the second figure, OliVe mislabels boat as a van, whereas MicroScopiQ is able to correctly identify it as a boat.

\subsection{CNN and SSM Quantization Results}
\label{sec:cnn_ssm}
In \autoref{tab:cnn_ssm_results}, we demonstrate MicroScopiQ quantization on CNNs and SSMs. MicroScopiQ achieves near lossless performance at W4A4, W4A8 configurations and $\leq 3\%$ accuracy drop at W2A4 setting. Similarly, MicroScopiQ achieves up to $30\%$ higher accuracy compared to the SoTA, QMamba \cite{li2025qmamba}, across SSM models.

\subsection{Accelerator Results}
\noindent \textbf{Compute Area. }In \autoref{tab:accelerator_area}, we compare the accelerator compute area breakdown of MicroScopiQ (w/ 1 ReCoN unit) with baselines GOBO and OliVe for a $64\times64$ array design. For a fair evaluation, all accelerators have identical configurations with same number of PEs. MicroScopiQ has a very low compute overhead $8.63\%$, compared to OliVe's $9.90\%$. This is because, units like ReCoN which perform three simple operations for outlier processing has a very low area utilization compared to OliVe's encoders and decoders. ReCoN is also time-multiplexed across all rows resulting in minimal area overhead (see \cref{sec:ablations}). Furthermore. the simple INT operations of MicroScopiQ allows the packing of more compute power with minimal overheads. A similar multi-precision design in OliVe would require 3.5-4$\times$ higher multi-precision support area due to complex exponent-integer PEs. GOBO has higher compute area compared to OliVe and MicroScopiQ due to large per-PE area.
\input{tables/accelerator_area}
\noindent \textbf{Maximum Performance Per Unit Area ($TOPS/mm^2$). }We also estimate the peak compute throughput per unit area of each accelerator (compute density), using LLaMA3-8B as the workload in \autoref{tab:accelerator_area}. To achieve peak throughput for MicroScopiQ, FMs must be quantized at $bb=2$. MicroScopiQ \textbf{achieves nearly 2$\times$ and 14$\times$ higher compute density} than OliVe and GOBO respectively.

\noindent \textbf{Iso-Accuracy Performance Comparison. }To enable a fair comparison of the latency of different accelerators, we perform an iso-accuracy comparison. We quantize different weight layers of the baseline models using $2$-, $4$-, or $8$-bit precisions, with activations at 4-bit (\autoref{fig:energy_perf}(a)). This ensures all model accuracies are within $\pm 2\%$ of the best quantized model i.e., W4A4 MicroScopiQ. In \autoref{fig:energy_perf}(b) we compare the normalized latency of baselines against two versions of MicroScopiQ, v1 (W4A4): all weight layers having $bb=4$ and v2 (WxA4): most layers quantized such that $bb=2$, with a small percentage of layers with $bb=4$ for iso-accuracy. MicroScopiQ v1 and v2 consistently outperform all baselines across models, achieving an \textbf{average speedup of 1.50${\times}$ and 2.47${\times}$} respectively. The higher speedup of MicroScopiQ v2 is due to most layers at $bb = 2$, allowing the PEs to perform higher throughput low precision computation, compared to $bb=4/8$.  

\noindent \textbf{Iso-Accuracy Energy Comparison. }In \autoref{fig:energy_perf}(c), we show the normalized energy consumption of different accelerators, composed of static and dynamic energy. MicroScopiQ v2 has the lowest energy consumption. Compared to designs like GOBO, OLAccel and AdaptivFloat, where computations happen at higher-precisions in 8-bit and 32-bit PEs, MicroScopiQ v2 has a significant advantage in terms of the core and DRAM dynamic energy consumption with simple INT PEs. On average MicroScopiQ v2, has \textbf{1.5$\times$ lower energy consumption} compared to baselines across different FMs.

\noindent 
\textbf{Power Breakdown. }The power breakdown of MicroScopiQ varies depending on the outlier distribution. For a LLaMA2-7B, the PE array consumes 56.23\% of power, followed by on-chip memory (36.80\%) and ReCoN (5.94\%). Conversely, VILA-7B, characterized by a higher outlier percentage, exhibits increased power consumption in the ReCoN unit (7.65\%) and slightly reduced on-chip memory (35.32\%), PE array (55.98\%) power consumption.

\begin{figure}[t]
    \centering\includegraphics[width=\columnwidth, keepaspectratio]{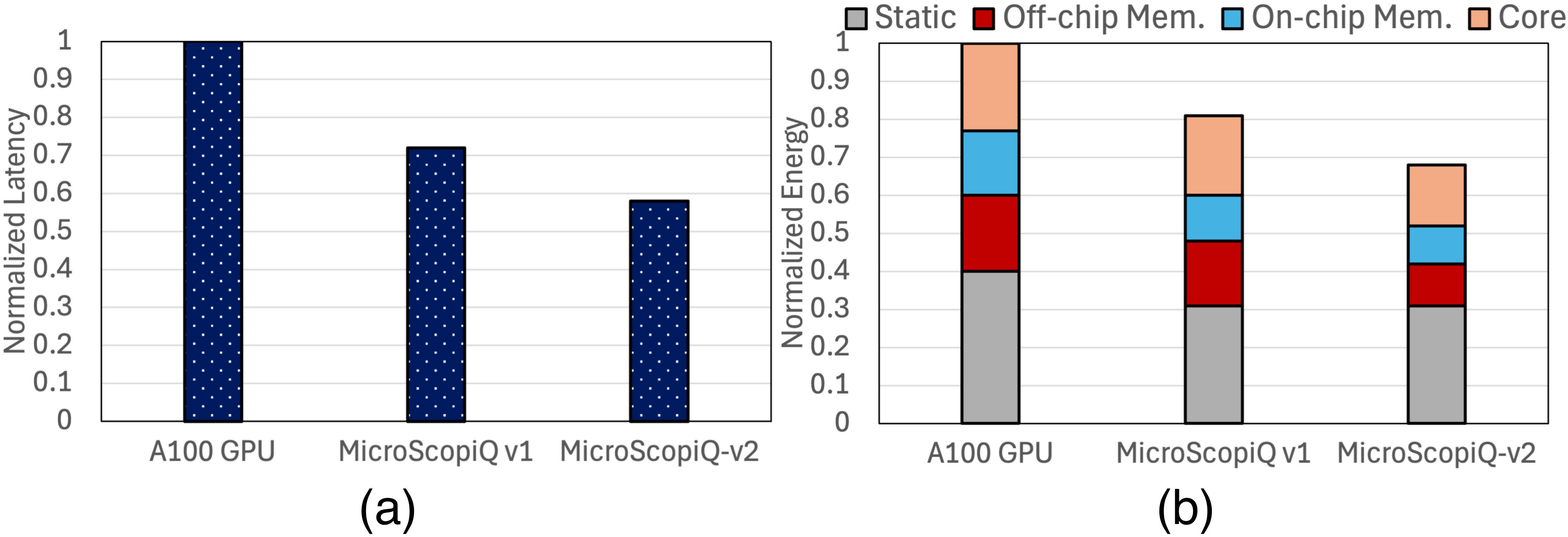}
        \vspace{-8mm}
        \caption{Comparison of MicroScopiQ accelerators with A100 GPU: (a) Normalized latency, (b) Normalized energy.} 

        \label{fig:gpu_accelerator}
        \vspace{-1mm}
\end{figure}

\vspace{-2mm}
\subsection{GPU Results}
\label{sec:gpu_results}
\textbf{Throughput on real GPU.} In \autoref{tab:quant_perf}, the unoptimized W4A4 MicroScopiQ (MS no-optim) underperforms the FP16 baseline due to the overhead of outlier merging in shared memory and GEMM execution in FP16. In contrast, our optimized W4A4 kernel (MS optim.) (\cref{sec:kernel}), achieves similar performance to SoTA technique Atom \cite{zhao2024atom} due to efficient register caching and dynamic GEMM acceleration, while simultaneously possessing better accuracy.

\noindent
\textbf{MicroScopiQ GPU simulation. } As shown in our simulated A100 GPU results (\autoref{tab:quant_perf}), our tensor core modification enables low-precision INT+FP GEMM, eliminating costly dequantization and FP16 GEMMs, achieving the highest throughput.

\input{tables/gpu_actual}

\noindent
\textbf{GPU v/s MicroScopiQ Accelerator. }We compare a \textbf{standard} A100 GPU with the MicroScopiQ accelerator under iso-bandwidth (off-chip: 2 TB/s, on-chip: 1.5kb/cycle/warp) \cite{raihan2019modeling} and iso-compute (55,296 multipliers \cite{NVIDIA_A100}) scenario. As shown in \autoref{fig:gpu_accelerator}(a), MicroScopiQ v1 (W4A4) and v2 (WxA4) (see above) yield 1.2$\times$ and 1.7$\times$ speedup, respectively, over A100 (W4A4), due to multi-precision PEs, avoiding FP16 computation. Additionally, ReCoN’s reordering is pipelined with partial sum reduction, significantly outperforming GPU \texttt{shfl\_sync}. Additionally, MicroScopiQ delivers superior energy efficiency (\autoref{fig:gpu_accelerator}(b)), whereas A100 incurs higher on-chip energy consumption, due to register-level reordering and FP16 overhead. \textit{Notably, traditional GPUs lack WmAn ($m \neq n$) acceleration, making MicroScopiQ accelerator an efficient alternative}.
\input{tables/ablation_2}
\input{tables/omni_microscopiq}
\vspace{-2mm}
\subsection{Algorithm ablations}
\label{sec:ablations}
\noindent
\textbf{Per-component accuracy impact. }In \autoref{tab:ablation_2} we examine the accuracy gained or lost by progressively incorporating the quantization techniques proposed in MicroScopiQ. We initially adopt INT-4 scalar quantization on the LLaMA3-8B full-precision baseline. We then apply MX-INT-4$_{128}$. However, on decreasing precision to 2-bit INTs i.e., MX-INT-2$_{128}$, we observe a spike in PPL, due to higher outlier quantization error. Upon quantizing outliers to MX-FP-4$_{8,8}$, we regain the lost performance due to preserving outlier values at higher-precision and FP-format. Furthermore, we observe that pruning of least important weights causes a minor increase in PPL, which is regained by Hessian update. We observe that quantizing activations with simple MX-INT-8$_{128}$ with a high $\alpha$ results in a very minimal increase in PPL due to the robustness of MicroScopiQ.   

\noindent
\textbf{KV-cache quantization. }In \autoref{tab:ablation_2} we also report the impact of KV-cache quantization on performance. Following the 2-bit KV-cache quantization technique proposed in \cite{liu2024kivi} we quantize K per channel, and the V per token with a MaB of size 128 and residual token length (R) \cite{liu2024kivi} of 128 for both key and value.

\noindent
\textbf{Omni-MicroScopiQ. }Our method is orthogonal to the techniques proposed in OmniQuant. We combine our method with OmniQuant to further improve the quantization performance of MicroScopiQ. We employ OmniQuant's Learnable Weight Clipping (LWC) to learn both the inlier and outlier scale factors of MicroScopiQ and leverage the Learnable Equivalent Transformation (LET) to learn to migrate the quantization difficulty of activations to weights. In \autoref{tab:microscopiq_omniquant} we compare the performance of MicroScopiQ enhanced with OmniQuant (Omni-MicroScopiQ) against OmniQuant. Omni-MicroScopiQ delivers improvements up to \textbf{22\%}.

\begin{figure}[t]
    \vspace{-2mm}
    \centering\includegraphics[width=0.85\linewidth, keepaspectratio]{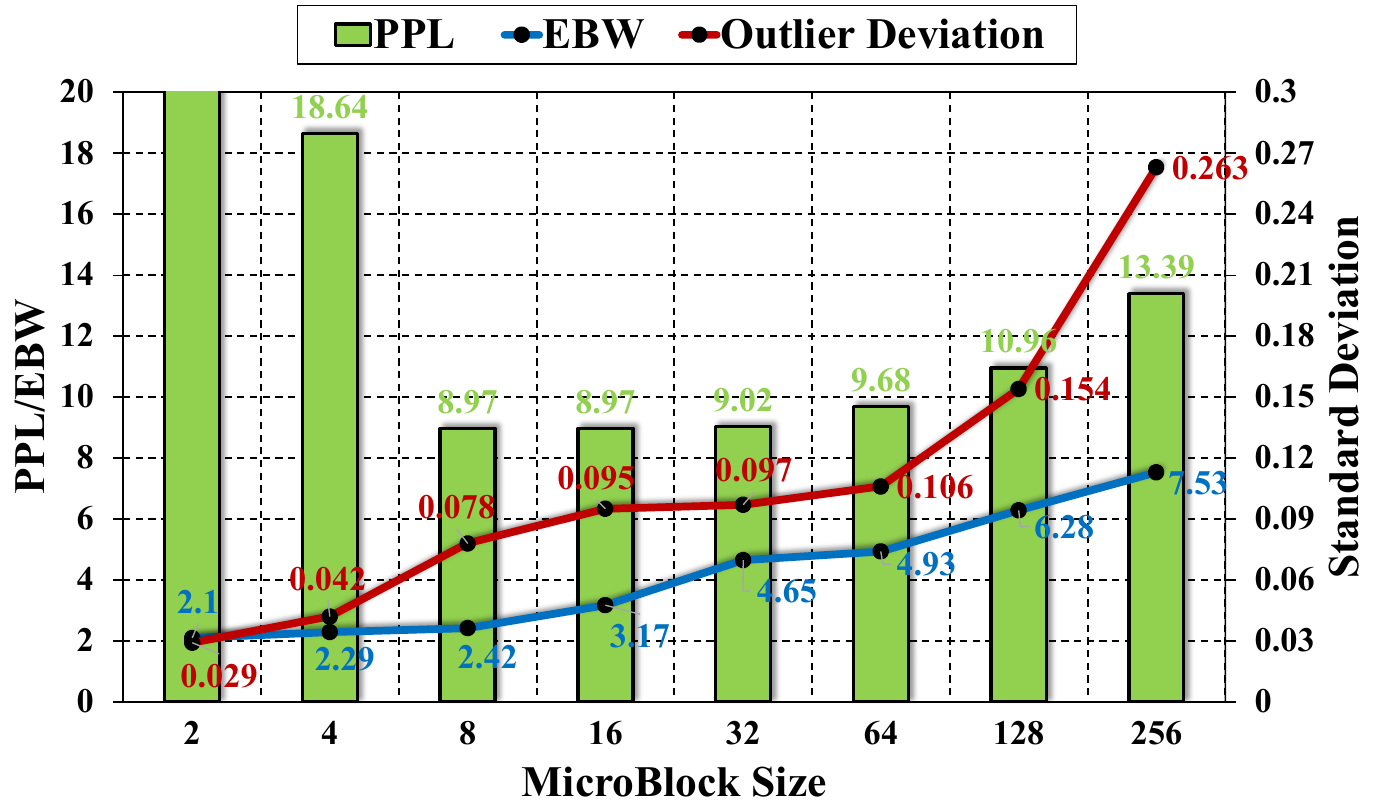}
        \vspace{-4mm}
        \caption{Effect of outlier group size on the Wikitext-v2 perplexity (PPL) and EBW of a MicroScopiQ quantized LLaMA3-8B. $\infty$ is per-tensor granularity.}
        \label{fig:outlier_group}
\end{figure}

\noindent
\textbf{Outlier group size.} In \autoref{fig:outlier_group}, we analyze the effect of different $B_\mu$ values on outlier diversity (measure by the standard deviation: \textcolor{red}{red line}), PPL (\textcolor[HTML]{32CD32}{green bar}), and EBW (\textcolor[HTML]{003152}{blue line}) for LLaMa3-8B. For $B_\mu=2,4$, accuracy degradation occurs due to outlier pruning, as many mBs have $\ge B_\mu/2$ outliers. As $B_\mu$ increases, outliers further apart show greater diversity, evidenced by the higher standard deviation. For $B_\mu \geq 32$, this results in increased quantization error and higher PPL, as MX-FP scales are shared across more diverse outlier values. Larger group sizes also significantly increase EBW due to metadata overhead. A balance is achieved at $B_\mu=8$.    


\begin{figure}
     \centering
        \includegraphics[width=\linewidth, keepaspectratio]{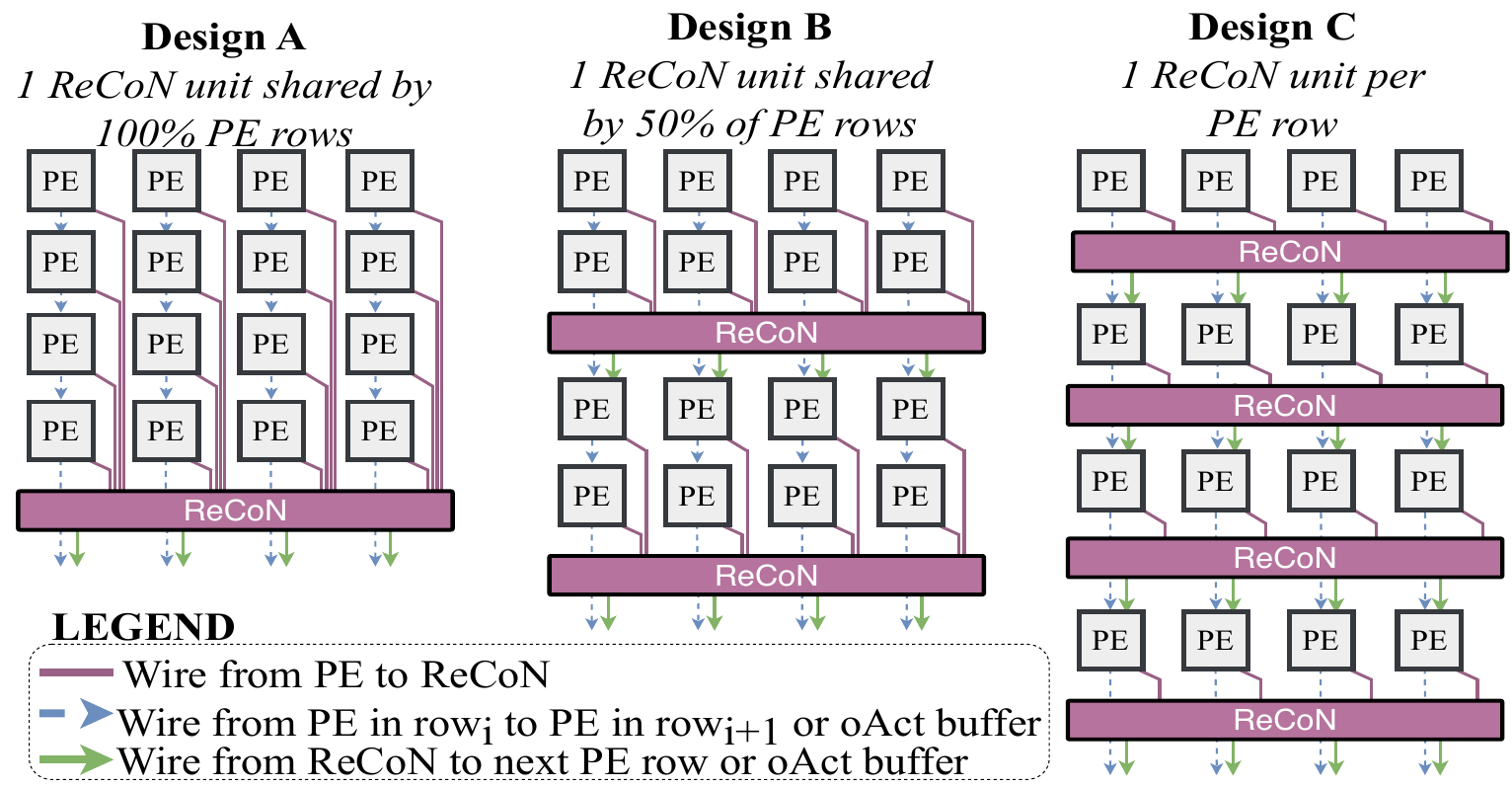}
        \vspace{-8mm}
        \caption{Different variants of MicroScopiQ accelerator with varying number of ReCoN units.}

        \label{fig:scale2}
\end{figure}

\subsection{Analysis of time-multiplexed ReCoN}
\label{sec:recon_analysis}
\noindent
\textbf{Accelerator design variants. }In \autoref{fig:scale2}, we show the layout of different variants of MicroScopiQ with varying number of ReCoN units, each of which can be employed based on application requirements. For area critical applications, design A, with a single ReCoN can be employed to have minimal area utilization with a slight increase in latency due to access conflicts to ReCoN. With the addition of more ReCoN units, fewer rows of PEs share a single ReCoN, dramatically reducing access conflicts to the shared ReCoN unit (design B). When the number of ReCoN units equals the number of rows, each PE row gets a dedicated ReCoN unit as shown in design C, which can be employed for latency critical applications. As we shall demonstrate subsequently, we identified that in up to a $128\times128$ PE array size, a total of 8 ReCoN units are sufficient to achieve peak performance with zero access conflicts to ReCoN.

\begin{figure}
     \centering
        \includegraphics[width=0.9\columnwidth, keepaspectratio]{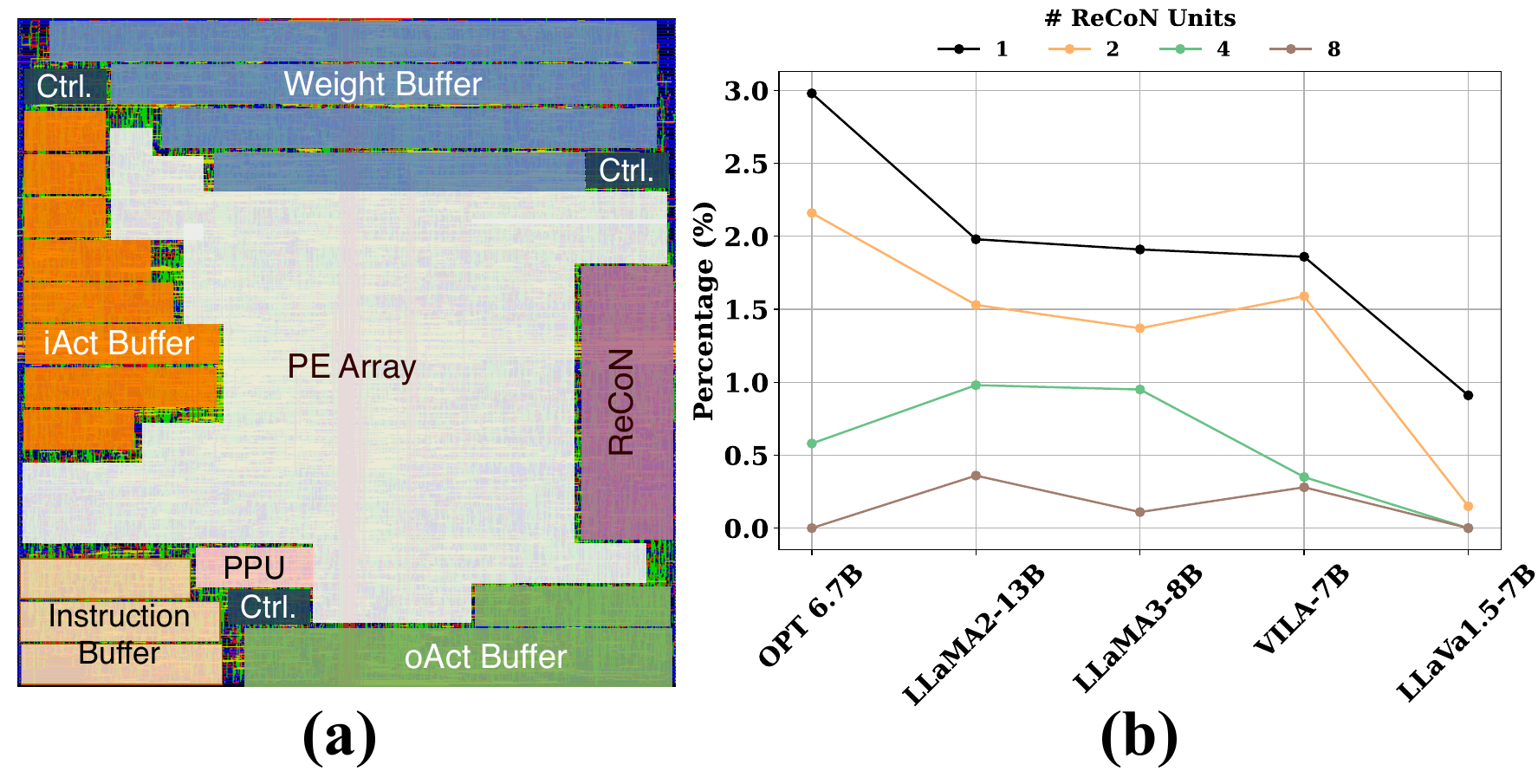}
        \vspace{-5mm}
        \caption{(a) Annotated MicroScopiQ floorplan ($64\times64$ array). (b) Percentage access conflicts to ReCoN for $64\times64$ PE array.}

        \label{fig:die_recon}
\end{figure}

\noindent
\textbf{Optimal \# of ReCoN units. }For a $64\times64$ MicroScopiQ accelerator, we study the number of access conflicts for different number of ReCoN units in \autoref{fig:die_recon}(b). Evidently, the number of access conflicts to ReCoN which is measured as the percent of total number of accesses to ReCoN that result in conflicts is under 3\%. With the progressive addition of more ReCoN units, the access conflicts tend to 0\%. As observed in \autoref{fig:energy_perf}, despite the access conflicts, 1 ReCoN unit is sufficient to reach optimal inference performance in an iso-accuracy scenario. For latency critical applications, in \autoref{fig:recon}(a), we quantify the performance improvement and impact on compute area by incorporating multiple ReCoN units for a LLaMA3-8B FM. Up to 8 ReCoN units enables peak performance with \textbf{21\% improved latency} and only \textbf{$1.58\times$ higher} \textbf{compute area}. For large array sizes, up to 8 ReCoN units are sufficient for peak performance.             

\subsection{Scalability and Timing Analysis}

\begin{figure}
     \centering
        \includegraphics[width=\columnwidth, keepaspectratio]{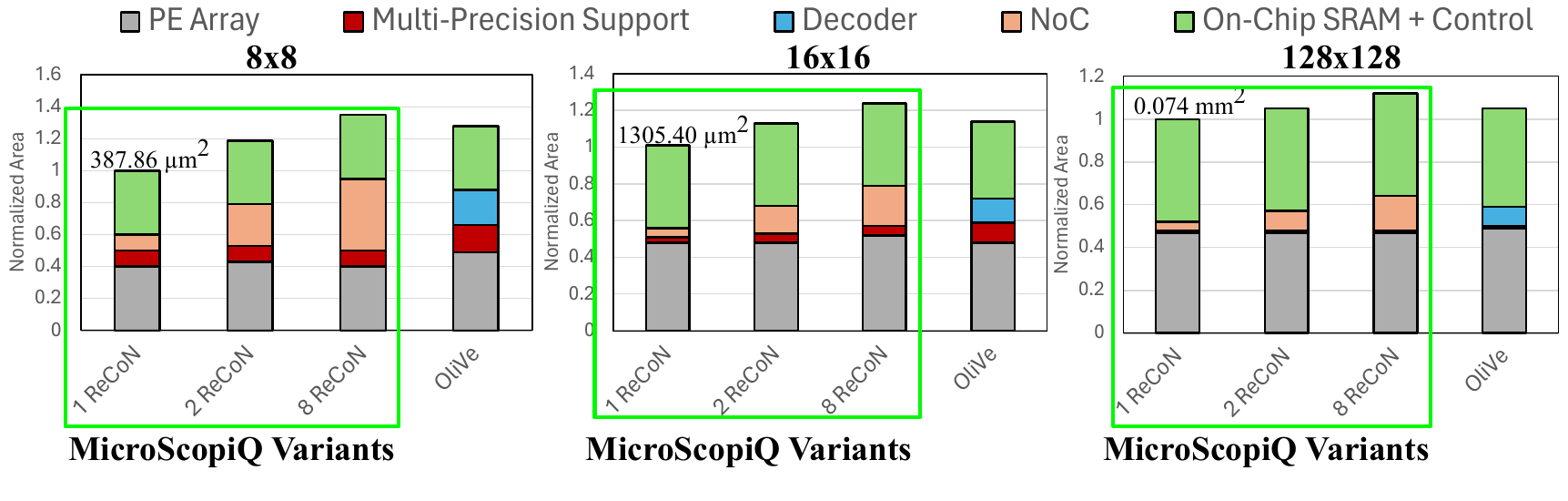}
        \vspace{-8mm}
        \caption{Comparison of area of MicroScopiQ and OliVe at different PE array sizes. Three different versions of MicroScopiQ are depicted with different number of ReCoN units.}
        \label{fig:scale3}
\end{figure}

\textbf{Scalability. }To study how MicroScopiQ scales to different PE array sizes, we compare the total area (including on-chip weight and iAct/oAct buffers) against OliVe at three different scales: $8\times8, 16\times16, 128\times128$ in \autoref{fig:scale3}. Following \cite{sumbul2022system}, for $8\times8$ array we employ 16 kB iAct and oAct buffers and 32 kB weight buffer. We also depict the accelerator floorplan for a $64\times 64$ array in \autoref{fig:die_recon}(a). We progressively scale the buffers for larger arrays to ensure full array utilization. Since varying the number of ReCoN units trades off area for performance, we also compare three versions of MicroScopiQ with 1, 2, 8 ReCoN units. For the single ReCoN variant of MicroScopiQ, as the design scales to higher array sizes, the area overhead of ReCoN decreases ($128\times128$ has overhead of 3\%), with the on-chip area being dominated by the large PE array and buffers. We observe similar trends for 2 or 8 ReCoN units. At large PE array sizes such as $128\times128$, adding 8 ReCoN units results in a minimal overhead of just 11\%. Particularly at large array sizes the column-wise bus from PEs to ReCoN can be routed over logic as the PE array scales. With same buffer configurations, OliVe has a higher area compared to a single ReCoN MicroScopiQ variant, while comparable area to MicroScopiQ with 8 ReCoN units. At all variants, MicroScopiQ has $2-3\times$ higher performance than OliVe (see \autoref{tab:accelerator_area}).           

\noindent
\textbf{Timing Analysis. }MicroScopiQ at all scales attains a peak clock frequency of 1 GHz. The critical path for MicroScopiQ is from the local weights registers to multipliers within the PE. The column-wise bus (the longest wire in MicroScopiQ) from PEs to ReCoN is not on the critical path and therefore does not limit scalability.
\noindent

\noindent
\textbf{Implementation overhead on NoC-based accelerators. }The global column-wise bus is a design commonly seen in AI ASICs like \cite{chen2019eyeriss, firoozshahian2023mtia, shao2019simba, tong2024feather}. These global buses are used in conjunction with the NoC in these designs for data distribution, layout reordering \cite{tong2024feather}. We modify two large-scale accelerators MTIA and Eyeriss-v2 that employ NoCs in their design to support MicroScopiQ, by incorporating ReCoN functionality in the accelerator NoCs and modifying PEs to support all MicroScopiQ specific operations. As shown in \autoref{fig:recon}(b), compared to the baseline design of MTIA and Eyeriss-v2, integrating ReCoN and MicroScopiQ PEs results in only a $3\%$ and $2.3\%$ increase in compute area, respectively. Since these designs already employ NoCs, integration overhead of MicroScopiQ is minimal. This demonstrates that the benefits achieved by MicroScopiQ are attained at minimal to no additional cost.

\vspace{-3mm}

%% file: tables/additional_llm_results.tex
\begin{table}[t]
    \centering
    \caption{Comparison of quantization performance for LLaMA2-70B on LLM benchmarks.}
    \vspace{-5pt}
    \resizebox{\linewidth}{!}{%
    \begin{tabular}{lc|cccc}
    \Xhline{2\arrayrulewidth}
    \rowcolor[HTML]{E0E0E0}
    \textbf{Method} & \textbf{W/A} & \textbf{ARC-c} & \textbf{HellaSwag} & \textbf{MMLU} & \textbf{WinoGrande} \\ 
    \rowcolor[HTML]{D9EAF5}
    Baseline & 16/16 & 60.50 & 84.30 & 68.90 & 80.60 \\
    \Xhline{1\arrayrulewidth}
    OliVe \cite{guo2023olive} & 2/16 & 38.60 & 55.30 & 39.80 & 60.70 \\
    OmniQuant \cite{shao2023omniquant} & 2/16 & 49.70 & 77.80 & 58.20 & 74.20 \\
    \rowcolor[HTML]{D5E8D4}
    \textbf{MicroScopiQ (Ours)} & \textbf{2/16} & \textbf{53.30} & \textbf{81.60} & \textbf{63.70} & \textbf{77.80} \\
    \Xhline{2\arrayrulewidth}
    \end{tabular}}
    \label{tab:llm_benchmarks}
\end{table}

%% file: tables/non_transformer.tex
\begin{table}[t]
    \centering
    \caption{Quantization results for non-transformer models. We report Top-1 accuracy (\%) on ImageNet (higher is better).}
    \vspace{-4.2mm}
    \resizebox{0.9\linewidth}{!}{%
    \begin{tabular}{lc|cc|cc}
    \Xhline{2\arrayrulewidth}
    \rowcolor[HTML]{E0E0E0}
    \textbf{Method} & \textbf{W/A} & \textbf{ResNet50} & \textbf{VGG16} & \textbf{VMamba-S} & \textbf{Vim-S} \\
    \Xhline{2\arrayrulewidth}
    \rowcolor[HTML]{D9EAF5}
    Baseline & 16/16 & 76.15\% & 71.59\% & 83.60\% & 80.50\% \\
    \Xhline{1\arrayrulewidth}
    \rowcolor[HTML]{FFFFFF}
    HAWQ \cite{dong2019hawq} & 2/4 & 73.17\% & 68.81\% & - & - \\
    \rowcolor[HTML]{FFFFFF}
    QMamba \cite{li2025qmamba} & 4/4 & - & - & 40.36\% & 68.50\% \\
    \rowcolor[HTML]{D5E8D4}
    \textbf{MicroScopiQ} & \textbf{4/4} & {75.08\%} & {70.84\%} & \textbf{70.07\%} & {71.52\%} \\
    \rowcolor[HTML]{D5E8D4}
    \textbf{MicroScopiQ} & \textbf{2/8} & \textbf{75.12\%} & \textbf{70.87\%} & {66.52\%} & \textbf{71.98\%} \\
    \rowcolor[HTML]{D5E8D4}
    \textbf{MicroScopiQ} & \textbf{2/4} & {73.61\%} & {69.12\%} & - & - \\
    \Xhline{2\arrayrulewidth}
    \end{tabular}}
    \label{tab:cnn_ssm_results}
\end{table}


%% file: tables/accelerator_area.tex
\begingroup	
\begin{table}[t]\centering
 \caption{Compute area and density comparison for a $64\times64$ array at 7nm process technology.}
 \vspace{-10pt}
 \renewcommand*{\arraystretch}{1.0}
  \setlength\tabcolsep{1.9pt}
\resizebox{\linewidth}{!}{%
\begin{tabular}{c|c|c|c|c}
\Xhline{2\arrayrulewidth}
 \makecell{Architecture} & \makecell{Component  \\ $\langle$ Area ($\mu m^2$), \# Units $\rangle$} & \makecell{Compute area \\ ($mm^2$)} & \makecell{Compute \\ overhead} & \makecell{Compute density \\ (TOPS/$mm^2$)} \\
  \Xhline{2\arrayrulewidth}
\multirow{3}{*}{GOBO \cite{zadeh2020gobo}} & Group PE $\langle 36.56, 4096\times \rangle$ & \multirow{3}{*}{0.216} & \multirow{3}{*}{3.28\%} & \multirow{3}{*}{28.28} \\
    \cline{2-2}

    & Outlier PE $\langle 96.42, 64\times \rangle$ &  & \\ \cline{2-2}
    & Control unit  $\langle 115.36, 1\times \rangle$ &  & \\ \cline{1-5}



\multirow{5}{*}{OliVe \cite{guo2023olive}} & 4-bit Decoder $\langle 1.86, 128\times \rangle$ & \multirow{5}{*}{{\textbf{0.011}}} & \multirow{5}{*}{{9.90\%}} & \multirow{5}{*}{{{184.30}}} \\
    \cline{2-2}

    & 8-bit decoder $\langle 2.47, 64\times \rangle$ &  & \\ \cline{2-2}
    & Base PE $\langle 2.51, 4096\times \rangle$ &  & \\ \cline{2-2}
    & Multi-Precision support $\langle 0.68, 1024\times \rangle$ &  & \\ \cline{2-2}
    & Control unit  $\langle 95.49, 1\times \rangle$ &  & \\ \cline{1-5}

\multirow{6}{*}{{\textbf{MicroScopiQ}}} & ReCoN $\langle 204.68, 1\times \rangle$ & \multirow{6}{*}{{0.012}} & \multirow{6}{*}{{{\textbf{8.63\%}}}} & \multirow{6}{*}{{\textbf{367.51}}} \\
    \cline{2-2}
    & Sync buffer $\langle 20.45, 1\times \rangle$ &  & \\ \cline{2-2}
    & Base PE $\langle 2.82, 4096\times \rangle$ &  & \\ \cline{2-2}
    & Multi-precision support $\langle 0.22, 4096\times \rangle$ &  & \\ \cline{2-2}
    & Control unit $\langle 105.78, 1\times \rangle$ &  & \\
\Xhline{2\arrayrulewidth}
\end{tabular}
}
\label{tab:accelerator_area}

\end{table}
\endgroup

%% file: tables/gpu_actual.tex
\begin{table}[t]
    \centering
    \caption{Normalized token generation throughput comparison across different quantization methods (Wikitext perplexity in brackets) on A100 GPU and GPU simulation \cite{bakhoda2009analyzing}.}
    \vspace{-4mm}
    \resizebox{\linewidth}{!}{%
    \begin{tabular}{l l | c c}
        \Xhline{2\arrayrulewidth}
        \rowcolor[HTML]{E0E0E0}
        \textbf{Device} & \textbf{Method} & \textbf{LLaMA-2 13B} & \textbf{LLaMA-3 8B} \\
        \Xhline{2\arrayrulewidth}
        \rowcolor[HTML]{D9EAF5}
        \multirow{4}{*}{\textbf{A100}} 
        & TRT-LLM FP16 & 1.00 (4.83) & 1.00 (6.13) \\
        & W4A4 Atom & 2.25 (6.12) & 1.05 (8.12) \\
        & \textbf{W4A4 MS$^*$ no-optim.} & 0.98 (5.57) & 0.92 (8.12) \\
        \rowcolor[HTML]{D5E8D4}
        & \textbf{W4A4 MS$^*$ optim.} & {2.06} (5.57) & {1.01} (8.12) \\
        \Xhline{1\arrayrulewidth}
        \rowcolor[HTML]{D5E8D4}
        \textbf{GPGPU Sim} & \textbf{W4A4 MS$^*$ w/ New MTC$^\ddagger$} & \textbf{4.31} (5.57) & \textbf{1.78} (8.12) \\
        \Xhline{2\arrayrulewidth}
            \rowcolor[HTML]{FFFFFF}
        \multicolumn{4}{l}{$^*$MS: MicroScopiQ; $^\ddagger$MTC: Modified Tensor Core}\\
        \Xhline{2\arrayrulewidth}
    \end{tabular}}
    \label{tab:quant_perf}
\end{table}

%% file: tables/ablation_2.tex
\begingroup	
\begin{table}[t]\centering
 \caption{Effect on Llama3-8B PPL \cite{meta2024introducing} upon progressive inclusion of different quantization techniques.}
 \vspace{-4mm}
\resizebox{\linewidth}{!}
{%
\begin{tabular}{ll}
\Xhline{2\arrayrulewidth}
\rowcolor[HTML]{E0E0E0}
\textbf{Quantization Method} & \textbf{WikiText2 PPL $\downarrow$} \\
\Xhline{1\arrayrulewidth}
\rowcolor[HTML]{D9EAF5}
Baseline W16A16 & 6.13 \textcolor[HTML]{A9A9A9}{(-)} \\
\rowcolor[HTML]{E0E0E0}
$+$ Quantize all weights to INT-4 & 10.27 \textcolor[HTML]{CC0000}{($\uparrow 4.14$)} \\
$+$ Quantize all weights to MX-INT-4$_{128}$ & 9.53 \textcolor[HTML]{009900}{($\downarrow 0.74$)} \\
\rowcolor[HTML]{E0E0E0}
$+$ Quantize all weights to MX-INT-2$_{128}$ & 39.48 \textcolor[HTML]{CC0000}{($\uparrow 29.95$)} \\
$+$ Quantize outliers to MX-FP-4$_{128,128}$ & 10.96 \textcolor[HTML]{009900}{($\downarrow 28.52$)} \\
\rowcolor[HTML]{E0E0E0}
$+$ Quantize outliers to MX-FP-4$_{8,8}$ & 8.93 \textcolor[HTML]{009900}{($\downarrow 2.03$)} \\
$+$ Reduce outlier mag. by $\times 2^{I_{sf}}$ & 8.89 \textcolor[HTML]{009900}{($\downarrow 0.04$)} \\
\rowcolor[HTML]{E0E0E0}
$+$ Prune least imp. inliers per $\mu B$ & 9.02 \textcolor[HTML]{CC0000}{($\uparrow 0.13$)}\\
$+$ Compensate quantization errors/rB & 8.97 \textcolor[HTML]{009900}{($\downarrow 0.05$)}\\
\rowcolor[HTML]{E0E0E0}
$+$ Quantize activations to MX-INT-8$_{128}$, $\alpha = 0.7$ &9.08 \textcolor[HTML]{CC0000}{($\uparrow 0.11$)}\\
$+$ 2-bit KV-cache quantization \cite{liu2024kivi} & 9.58 \textcolor[HTML]{CC0000}{($\uparrow 0.50$)}\\
\Xhline{2\arrayrulewidth}
\end{tabular}
}

\label{tab:ablation_2}
\end{table}
\endgroup

%% file: tables/omni_microscopiq.tex
\begin{table}[t]
    \centering
    \caption{LLM quantization performance of MicroScopiQ integrated with OmniQuant.}
    \vspace{-4mm}
    \resizebox{\linewidth}{!}{%
    \begin{tabular}{lc|ccc}
    \Xhline{2\arrayrulewidth}
    \rowcolor[HTML]{E0E0E0}
    \textbf{Method} & \textbf{W/A} & \textbf{Llama-2 13B} & \textbf{Llama-3 70B} & \textbf{Phi-3 3.8B} \\ 
    \rowcolor[HTML]{D9EAF5}
    Baseline & 16/16 & 4.83 & 2.85 & 6.33 \\
    \Xhline{1\arrayrulewidth}
    OmniQuant \cite{shao2023omniquant} & 4/16 & 5.02 & 3.46 & 6.67 \\
    \rowcolor[HTML]{D5E8D4}
    \textbf{Omni-MicroScopiQ} & \textbf{4/16} & \textbf{4.87} & \textbf{2.97} & \textbf{6.52} \\
    \Xhline{1\arrayrulewidth}
    OmniQuant \cite{shao2023omniquant} & 2/16 & 7.56 & 6.17 & 7.09 \\
    \rowcolor[HTML]{D5E8D4}
    \textbf{Omni-MicroScopiQ} & \textbf{2/16} & \textbf{6.58} & \textbf{5.09} & \textbf{6.89} \\
    \Xhline{1\arrayrulewidth}
    OmniQuant \cite{shao2023omniquant} & 2/8 & 8.92 & 6.83 & 7.95 \\
    \rowcolor[HTML]{D5E8D4}
    \textbf{Omni-MicroScopiQ} & \textbf{2/8} & \textbf{7.12} & \textbf{5.74} & \textbf{7.21} \\
    \Xhline{2\arrayrulewidth}
    \end{tabular}}
    \label{tab:microscopiq_omniquant}
    \vspace{-1mm}
\end{table}

%% file: sections/related_work.tex
\section{Related Work}
\noindent
\textbf{Model compression for LLMs. } LLM weight compression schemes like quantization often rely on mixed-precision quantization with different outlier and inlier bit-widths \cite{lee2024owq, dettmers2023spqr, zhao2024atom, jeong2024sdq, huang2024good}. Other class of quantization rely on channel-wise shared scaling with different scale factors for outliers \cite{lee2024owq, shao2023omniquant, huang2024good}. Other quantization techniques mitigate outliers from weights and activations \cite{liu2024spinquant, xiao2023smoothquant, ashkboos2024quarot}.

LLM weight pruning methods \cite{frantar2023sparsegpt, yin2023outlier, ashkboos2024slicegpt, ramachandran2025acceleratingllminferenceflexible} rely on approximate solvers to identify the importance of weights based on magnitude of weights or weights activation combined \cite{sun2023simple}. However, they fail to achieve satisfactory performance even at low sparsity levels\cite{lu2024spp}. \textit{MicroScopiQ provides a unified solution for outlier-aware quantization by enabling flexible pruning of least-important weights to distribute additional outlier bits, thereby, simultaneously achieving high accuracy and hardware efficiency.} 

\noindent
\textbf{Unified pruning and quantization. }SDQ~\cite{jeong2024sdq} combines pruning and quantization by decomposing weights into inlier and outlier vectors, each employing a fixed N:M pattern and different bit precisions. In contrast, MicroScopiQ removes the rigid N:M constraint and achieves hardware efficiency through a consistent bit-budget and unified data type per element. Moreover, unlike SDQ’s sequential pruning-then-quantization approach, \textit{MicroScopiQ integrates pruning implicitly within quantization, enabling tighter coupling.}

\noindent
\textbf{Accelerator for quantized models.} Accelerators like GOBO \cite{zadeh2020gobo}, OLAccel \cite{park2018energy} use high-precision PEs for outliers and low-precision PEs for inliers. OliVe \cite{guo2023olive} incorporates encoding/decoding units in a PE array for inlier-outlier formats, while other accelerators \cite{tambe2020algorithm, ramachandran2024algorithm, guo2022ant} propose adaptive formats and hybrid FP PEs. However, they often suffer from unaligned memory access and large overheads. \textit{MicroScopiQ accelerator introduces a novel low overhead NoC called ReCoN, that effectively abstracts FP-format complexity from INT-PEs} 

\noindent
\textbf{NoCs in ML Accelerators.} NoCs are widely used in accelerators: \cite{qin2020sigma, munoz2023flexagon} employ a Benes NoC for data distribution, \cite{tong2024feather} uses a butterfly NoC for layout transformation, and \cite{firoozshahian2023mtia, chen2019eyeriss} utilize mesh NoCs. \textit{ReCoN is a butterfly NoC employed to compute FP partial sums.}

\begin{figure}[t]
    \centering
    \includegraphics[width=\linewidth, keepaspectratio]{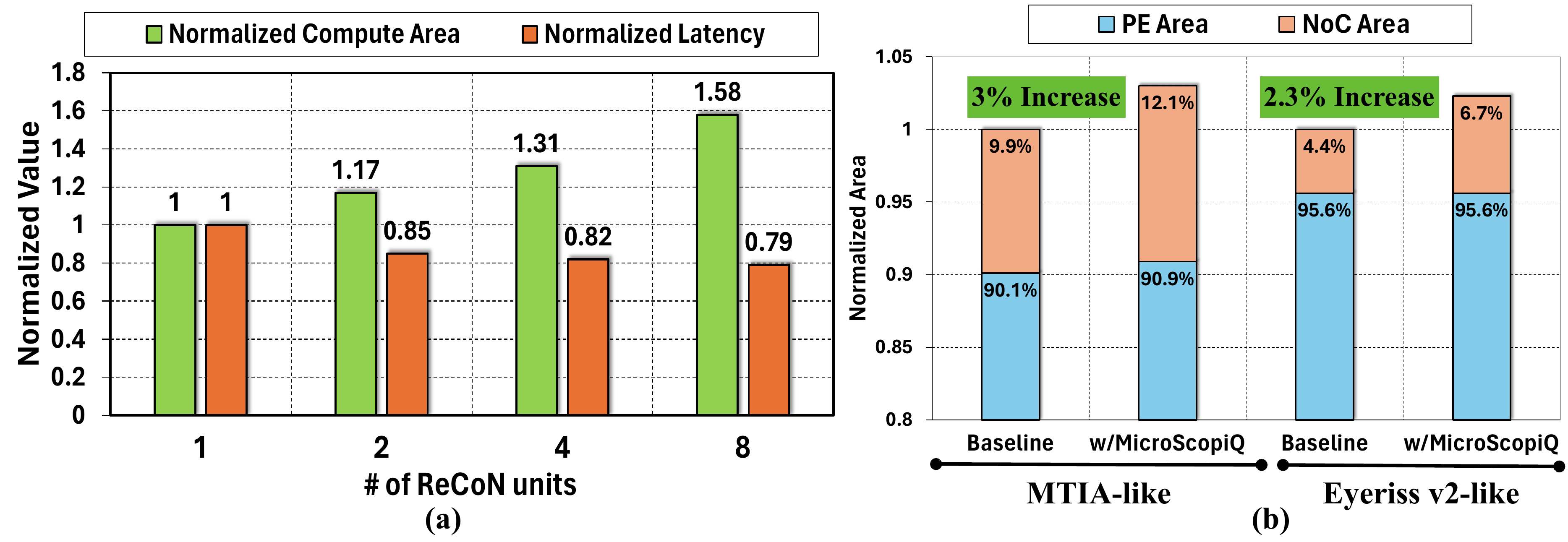}
        \vspace{-7mm}
        \caption{(a) Effect of time-multiplexed ReCoN units on the compute area and inference latency, (b) Implementation overhead of MicroScopiQ in NoC-based accelerators \cite{firoozshahian2023mtia, chen2019eyeriss}.}
        \label{fig:recon}

\end{figure}

\vspace{-3mm}